\documentclass[a4paper,fleqn]{cas-dc}

\usepackage[authoryear]{natbib}
\usepackage[abs]{overpic}
\usepackage{xcolor}
\usepackage{tabularx}
\usepackage{pict2e}
\usepackage[normalem]{ulem}

\usepackage{tikz}
\newcommand*\circled[1]{\tikz[baseline=(char.base)]{
            \node[shape=circle,draw,inner sep=2pt,very thick] (char) {#1};}}
\def\tsc#1{\csdef{#1}{\textsc{\lowercase{#1}}\xspace}}
\tsc{WGM}
\tsc{QE}
\tsc{EP}
\tsc{PMS}
\tsc{BEC}
\tsc{DE}
\begin{document}
\let\WriteBookmarks\relax
\def\floatpagepagefraction{1}
\def\textpagefraction{.001}

\shorttitle{An end-to-end KNN-based PTV approach for high-resolution measurements and uncertainty quantification}

\shortauthors{I. Tirelli et~al.}

\title [mode = title]{An end-to-end KNN-based PTV approach for high-resolution measurements and uncertainty quantification}                      
%\tnotemark[1,2]

%\tnotetext[1]{This document is the results of the research
 %  project funded by the National Science Foundation.}

%\tnotetext[2]{The second title footnote which is a longer text matter
 %  to fill through the whole text width and overflow into
  % another line in the footnotes area of the first page.}

\author{Iacopo Tirelli}[%type=editor,
                        %auid=000,bioid=1,
                        %prefix=Sir,
                        %role=Researcher,
                        orcid=0000-0001-7623-1161]

\cormark[1]


 %Email id of the first author
\ead{iacopo.tirelli@uc3m.es}

%\ead[url]{www.cvr.cc, cvr@sayahna.org}

%\credit{Conceptualization of this study, Methodology, Software}

\affiliation{organization={Department of Aerospace Engineering, Universidad Carlos III de Madrid},%Department and Organization
            addressline={Av. de la Universidad 30}, 
            city={Leganés},
            postcode={28911}, 
            country={Spain}}

\credit{Methodology, Software, Validation, Formal analysis, Investigation, Data Curation, Writing - Original Draft, Visualisation}

\author{Andrea Ianiro}[orcid=0000-0001-7342-4814]
\credit{Methodology, Investigation, Writing - Original draft, Writing - Review \& Editing, Supervision}

\author{Stefano Discetti}[orcid=0000-0001-9025-1505]
\credit{Conceptualisation, Methodology, Software, Investigation, Resources, Writing - Original draft preparation, Writing - Review \& Editing, Supervision, Project Administration, Funding Acquisition}

\cortext[cor1]{Corresponding author}
%\cortext[cor2]{Principal corresponding author}

%\fntext[fn1]{This is the first author footnote. but is common to third
%  author as well.}
%\fntext[fn2]{Another author footnote, this is a very long footnote and
  %it should be a really long footnote. But this footnote is not yet
 % sufficiently long enough to make two lines of footnote text.}

%\nonumnote{This note has no numbers. In this work we demonstrate $a_b$
 % the formation Y\_1 of a new type of polariton on the interface
%  between a cuprous oxide slab and a polystyrene micro-sphere placed
%  on the slab.
%  }

\begin{abstract}
We introduce a novel end-to-end approach to improving the resolution of Particle Image Velocimetry (PIV) measurements. We conceptualised the algorithm as a tool that is able, starting from raw pictures, to obtain high-resolution flow fields and uncertainty estimations with minimal intervention from the user. 
The method blends information from different snapshots without the need for time-resolved measurements on grounds of similarity of flow regions in different snapshots. 
The main hypothesis is that, with a sufficiently large ensemble of statistically-independent snapshots, the identification of flow structures that are morphologically similar but occurring at different time instants is feasible. 
Since the particles randomly seed the flow, a randomised sampling of such structures is naturally achieved, providing different views of the same region. 
Measured individual vectors from different snapshots with similar flow organisation can thus be merged, resulting in an artificially increased particle concentration. 
This allows refining the interrogation region and, consequently, increasing the spatial resolution. 
The measurement domain is split in subdomains. The similarity is enforced only on a local scale, i.e. morphologically-similar regions are sought only among subdomains corresponding to the same flow region. The identification of locally-similar snapshots is based on  $K$-nearest neighbours search in a space of significant flow features.
Such features are defined in terms of a Proper Orthogonal Decomposition, performed in subdomains on the original low-resolution data, obtained either with standard cross-correlation or with binning of Particle Tracking Velocimetry data with a relatively large bin size. 
A refined bin size is then selected according to the number of ``sufficiently close'' snapshots identified. 
The more neighbours identified, the higher the ``virtual'' particle image density and the smaller is the bin size, provided that the number of particles to be contained in it is fixed. 
The statistical dispersion of the velocity vectors within the bin is then used to estimate the uncertainty and to select the optimal $K$ which minimises it. 
The method is tested and validated against datasets with a progressively increasing level of complexity: two virtual experiments based on numerical simulations of the wake of a fluidic pinball and a channel flow and the experimental data collected in a turbulent boundary layer. 
\end{abstract}

% \includegraphics{figs/grabs.pdf}

%\item Research highlights item 1


% Keywords

\maketitle

\section{Introduction}
\label{sec:intro}
The characterisation of turbulent flows poses an exceptional challenge for measurement techniques. Turbulence dynamics involve spatial and temporal scales spanning a range whose extent increases with the Reynolds number. 
Particle Image Velocimetry \citep[PIV,][]{raffel1998particle}  uses seeding particles as  ``flow-samplers'', and evaluates the velocity from particle displacements between at least two snapshots with known time separation, recorded by a camera. Several image processing techniques can be used to extract flow quantitative descriptions in two or three dimensions \citep{raffel1998particle,westerweel2013particle,discetti2018volumetric}. PIV stands out for providing a spatial (and temporal, if sufficiently fast hardware is available) description of the organisation of turbulent flows, thus providing useful information for diagnostics, as well as for validation of models and numerical simulations, among others.

Since PIV is based on particles imaged on cameras as flow samplers, the range of observable scales depends directly on hardware limitations and mean particle spacing. 
The ratio between the largest and the smallest measurable scales is referred to as Dynamic Spatial Range (DSR), as reported by \citet{adrian1997dynamic}.
The largest measurable scale can be increased by employing larger camera sensors, and/or increasing the field of view by reducing the optical magnification. 
The smallest one depends on the capability of the particles to sample the flow field, i.e. on the particle concentration. PIV requires distinguishable particle images, thus setting an upper limit on the particle image density, i.e. the number of particles that can be imaged per unit area on the camera.
Similarly, a Dynamic Velocity Range (DVR) can be defined \citep{adrian1997dynamic} as the ratio between the largest and smallest measurable velocity. In state-of-art processing algorithms, the largest particle image displacement is limited mostly by the need of maintaining acceptable levels of in-plane velocity gradients (and of out-of-plane motion for planar PIV) while the smallest one depends on the accuracy of the process.

As the Reynolds number increases, turbulent flows impose progressively stricter constraints on the DSR and DVR needed for their complete characterisation. 
\citet{westerweel2013particle} demonstrated that the product between DSR and DVR to achieve measurements covering the span between large scales and the Kolmogorov scale should be of the same order of magnitude as the turbulent Reynolds number. 
The requirements, nonetheless, are much stricter on the DSR, since the ratio between the large scales and the Kolmogorov spatial scale broadens with $Re^{3/4}$, while for velocity the scaling follows $Re^{1/4}$ \citep{pope2000turbulent}. 
This has fostered over the years intense research to develop high-accuracy high-spatial-resolution PIV processing techniques, including cross-correlation-based multi-step image deformation methods \citep{scarano2001iterative} with  weighting windows \citep{nogueira1999local,astarita2007analysis}, adaptive-resolution techniques \citep{di2002windowing,theunissen2006adaptive,astarita2009adaptive,novara2012adaptive}, or methods exploiting time coherence in time-resolved measurements \citep{hain2007fundamentals,sciacchitano2012multi,cierpka2013higher,lynch2013high,schanz2016shake,beresh2021time}.

The above-mentioned methods target at improving the accuracy and spatial resolution of the interrogation process when dealing with individual PIV snapshots or straddles of time-resolved fields. It is though common practice in PIV experiments to capture large sequences of samples, most often statistically-independent from each other. 
This paves the way to improving the spatial resolution and the measurement accuracy by employing statistical information. 
This concept has been leveraged in the past for the extraction of high-resolution statistics (well beyond the DSR and DVR limitation of individual measurements). 
Remarkable examples include ensemble-correlation, also called single-pixel correlation \citep{westerweel2004single,scharnowski2012reynolds,avallone2015convergence}, and ensemble-particle-averaging \citep{cowen1997hybrid,kahler2012resolution,aguera2016ensemble}, often referred to as Ensemble Particle Tracking Velocimetry (EPTV). 
EPTV aims at obtaining dense clouds of vectors by superposition of instantaneous realisations. 
Bin averaging of such clouds delivers local probability distribution functions (PDF). 
Once the number of particles needed for acceptable convergence of the PDF is fixed, increasing the number of snapshots allows reducing progressively the bin size, thus increasing the spatial resolution.  
In recent years, most of the attention has been devoted to this second pathway, leaving slightly aside the development of ensemble correlation. 
The reason is threefold. 
First, Lagrangian Particle Tracking has proved to be superior to cross-correlation-based approaches in 3D \citep{kahler2016main}. Second, \citet{kahler2012resolution} demonstrated that the resolution limit of ensemble-correlation methods is the particle image diameter, thus making EPTV more suitable for regions where resolution is critical, e.g. near-wall measurements \citep{kahler2012uncertainty,sanmiguel2017adverse}. 
Third, the extraction of the statistical moments is significantly simpler and does not require any a priori assumption on the PDF shape of the fluctuations. 

Ensemble correlation and EPTV sacrifice the instantaneous information to obtain high-resolution statistics. 
Both methods ground on the hypothesis that individual realisations are samples of the same local statistical distribution. Ensemble correlation and EPTV leverage this principle to extract statistical moments. The path of blending information between samples that occur to be ``close'' within the statistical distribution, on the other hand, has been barely explored. 
Recently \cite{cortina2021sparse} proposed a Data-Enhanced PTV (DEPTV), based on EPTV to obtain high-resolution Proper Orthogonal Decomposition modes \citep[POD, ][]{lumley1967structure}. 
This process is equivalent to a Linear Stochastic Estimation of the original low-resolution PIV measurements on the temporal POD modes. 
DEPTV demonstrated being a suitable tool to enhance the spatial resolution, especially for cases with relatively compact POD eigenspectrum. 

The recent advances in data-driven and machine-learning algorithms have similarly stimulated the development of resolution-enhancement methods on grounds of analysis of the statistical distribution of the available samples. 
Optical flow estimators based on deep learning techniques \citep{cai2019dense,lagemann2021deep} have shown promising results, although their robustness and generalisability are still under investigation.
Successful recent super-resolution algorithms are based on Generative Adversarial Networks \citep[GANs,][]{goodfellow2014generative}. \citet{deng2019super} achieved an increase in spatial resolution of up to $8$ times with super-resolution GANs \citep{ledig2017photo}. \citet{guemes2022super} recently introduced a new concept of Randomly Seeded super-resolution GANs (RaSeedGAN) that achieves similar resolution enhancement and does not need a paired low-high resolution dataset for training as it exploits directly the sparse particle measurements as a target.The main drawback of neural-network-based methods is that they require experienced users for training, and the uncertainty quantification is still difficult, as we can see in the novel methodologies proposed by \citet{wang2022dense,wang2022deep} and \citet{gao2021robust}.

As discussed in the previous section, PIV is now a consolidated technique in fluid dynamic research. Nevertheless, all measurement techniques are affected by uncertainty sources, and PIV is not an exception. Uncertainty quantification in PIV has received significant attention in the last decade and it is now recognised as a fundamental step of the measurement chain. 

In this work we propose a novel methodology to blend information from different snapshots, and directly embed uncertainty quantification in the process. We divide the measurement domain in subdomains. For each snapshot, we search neighbours of each subdomain to merge their corresponding information. The method can be considered a local ensemble particle averaging, which is performed on each region only among snapshots which are identified as neighbours in the statistical distribution. The similarity between subdomains in different snapshots is assessed in terms of a $K$-nearest neighbour search in a reduced-dimensionality space. This feature space is obtained by performing a POD analysis in each subdomain separately, i.e a ``local'' POD.
The statistical dispersion among the particles identified in the nearest neighbours, and used for averaging, is exploited to provide an estimation of the measurement uncertainty. 

All these steps are carried out without any intervention from the user. In this way, it can be executed as a closed tool that is able to build high-resolution flow fields from the raw images and provide, in addition, a precise uncertainty estimation. On the other hand, a user with a minimum of expertise can refine the process parameters to adjust the output.

The details of the proposed method are discussed in Sec.~\ref{sec: algorithm description}, where all the steps are highlighted and their theoretical background is explained. 
In order to analyse the algorithm performance, in Sec.~\ref{sec: numerical validation} it is tested with two different datasets with increasing complexity, the flow around the fluidic pinball and the flow in a turbulent channel. In Sec.~\ref{sec: exp validation} the proposed methodology is also validated on an experimental dataset.

\section{Methodology} \label{sec: algorithm description}

\begin{figure*}[htb]
\centering 
\begin{overpic}[scale=1,unit=1mm]{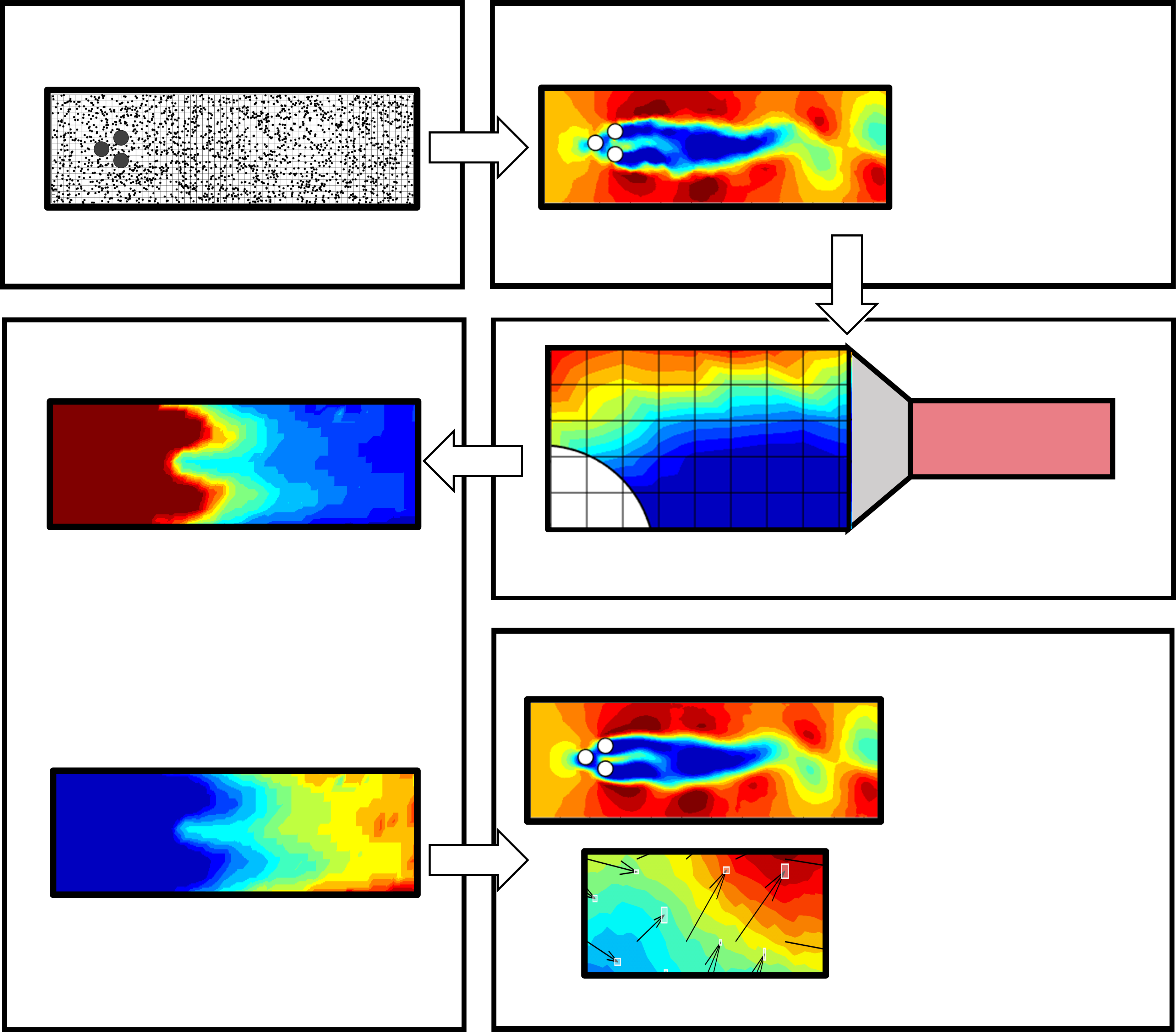}
\put(0,118){\parbox{55mm}{\centering \textnormal{\textbf{1. PTV data}}}}
\put(0,80){\parbox{55mm}{\centering \textnormal{\textbf{4.a Optimal $K$ map }}}}
\put(0,45){\parbox{55mm}{\centering $1< K_{op} < K_{max}$ \\ \textnormal{$K_{op}$ is the one  that} \\ \textnormal{ minimizes  the uncertainty}}}
\put(0,33){\parbox{55mm}{\centering \textnormal{\textbf{4.b Target bin size map }}}}
\put(0,8){\parbox{55mm}{\centering $b_{HR} = \sqrt{\frac{N_{p}}{N_{ppp}K}}$}}
\put(59,118){\parbox{81mm}{\centering {\textnormal{\textbf{2. Reference binned distribution }}}}}
\put(110,80){\parbox{30mm}{\centering \textnormal{\textbf{3. Local POD }} }}
\put(110,105){\parbox{30mm}{\centering $b_{ref} = \sqrt{\frac{N_{p}}{N_{ppp}}}$ \\  \centering $c_{i} = e ^ {-\left(\frac{d_{i}}{b_{ref}/2}\right)^2}$}}
\put(59,42){\parbox{81mm}{\centering \textnormal{\textbf{5. KNN-PTV reconstruction \& uncertainty }}}}
\put(108,18){\parbox{40mm}{\begin{flushleft}
 $ \hat{\delta} = \sqrt{\frac{\sum\limits_{i = 1}^{N_p} c_i\left(\boldsymbol{u_{p_{i}}}-\frac{\sum\limits_{i = 1}^{N_p} {(c_i\boldsymbol{u_{p_{i}}}})}{\sum\limits_{i=1}^{N_p}{c_i}}\right)^2}{\frac{M-1}{M} \sum\limits_{i = 1}^{N_p} c_i}}$ \\ $ c_{i} = e ^ {-\left(\frac{d_{i}}{b_{HR}/2}\right)^2}e ^ {-\left(4d_{f_i}\right)^2}$ \\ $ \boldsymbol{u}(x,y) = \frac{\sum\limits_{i = 1}^{N_p}{\boldsymbol{u_{p_{i}}} c_{i}} }{\sum\limits_{i = 1}^{N_p}{c_i}}$
\end{flushleft} }}
\put(109,70){\parbox{23mm}{\centering\textnormal{Training Set}\\$\boldsymbol{\Psi_{w}\Sigma_{w}}$}}
\put(66,55){\parbox{35mm}{\centering $ \boldsymbol{U_{w}} =  \boldsymbol{\Psi_{w}\Sigma_{w}\Phi_{w}^T}$}}
\end{overpic}
\caption{Flowchart of KNN-PTV. }
\label{Fig.flow}
\centering
\end{figure*}
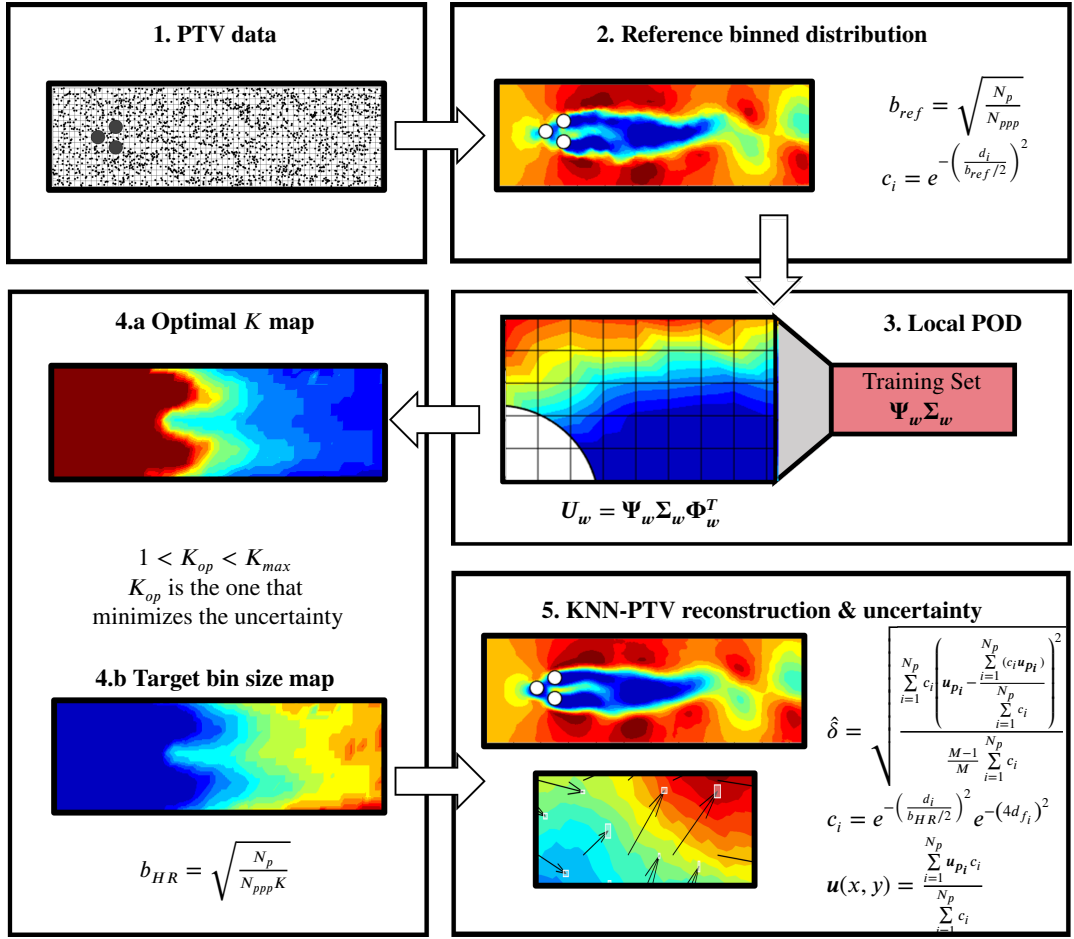

The flowchart of the proposed algorithm for high-resolution field reconstruction is sketched in Fig.~\ref{Fig.flow}. The method builds upon two hypotheses: 
\begin{itemize}
    \item with a sufficiently large ensemble of velocity fields, the identification of subdomains that are morphologically similar in different snapshots is feasible;
    \item the particles randomly sample the flow velocities in statistically-independent snapshots. This provides randomised sampling of flow structures which are assessed to be similar, even if occurring at different time instants in a fixed location.
\end{itemize} 
The similarity between realisations in the same local flow region is assessed in a low-order feature space. Such space is built by performing a POD analysis on each subdomain separately, and using the most energetic local POD modes. We refer to this process as ``local POD'' in the remainder of the paper. A $K$-nearest neighbour (KNN) algorithm pinpoints the  ``closest'' candidates in the ensemble based on the coordinates described by POD. KNN  is a non-parametric supervised learning method developed by \citet{fix1989discriminatory}. The method we propose is referred from now on as KNN-PTV.

The steps of the method are outlined in the following, and summarised in Tab.~\ref{tab:algorithm}.

\vspace{0.1cm}
%  all the details will be explained in this section.
\textbf{Step 1: PTV analysis of original images} 

KNN-PTV is fed by individual scattered vectors from PTV analysis performed following the super-resolution approach proposed by \citet{keane1995super}. It must be remarked that more advanced methods can be used for particle matching, with expected improvement, especially on dense particle images. However, it is expected that KNN-PTV can only benefit from a more accurate input. In the end, we thus opted for the above mentioned one, which is robust enough for moderate particle image density, and leave the implementation with more advanced particle matching methods for future investigation. Nonetheless, a slightly lower particle image density if compared to standard PIV is advisable to reduce the occurrence of spurious matchings and/or overlapping particles.

\vspace{0.1cm}
\textbf{Step 2: Building reference binned distribution} 

Feature identification requires a set of input fields for local morphological analysis. This initial complete ``low-resolution distribution'' can either be obtained with standard PIV algorithms or binning the PTV data from Step 1 with a sufficiently large bin size. In this work, we follow the second approach to reconstruct the reference binned distribution on a Cartesian grid $\boldsymbol{x}=(x,y) \in \mathbb{R}^{n_x \times n_y}$, where $n_x$ and $n_y$ are respectively the numbers of grid points along the streamwise and axial direction.

The bin size must be selected in order to minimise the empty spots. In case of occurrence of empty spots, the value of velocity corresponding to them can be obtained by interpolation. Assuming that the binning is a simple moving-average operation, in order to obtain $N_p$ particles per spot it should be set a bin size (assumed square) such that:
\begin{equation}
   b_{ref} = \sqrt{\frac{N_{p}}{N_{ppp}}},
   \label{Eq.bref}
\end{equation}
%The obtained vectors are reduced to a cartesian grid to  build a reference for the search of similar snapshots with KNN. To this purpose, the domain is discretized into several  square bins and the particles falling inside each of them  are identified. 
%It must be noted here that the number of particles falling within a bin $N_p$ and the bin size are directly proportional. Since for each bin we estimate an average velocity, in this work, in order to obtain a successful convergence of the averaging process, $N_{p}$ is set to 10, although other choices are suitable depending on the user needs. The bin side length in pixels is then chosen to be equal to:
\noindent where $N_{ppp}$ is the particle density, expressed in particles per pixel, and $b_{ref}$ is the bin size of the reference binned fields. A Poisson distribution models quite accurately the probability of finding empty bins once $N_p$ is selected:

\begin{equation}
   Pr\{0\} = e^{-N_p}. 
   \label{Eq.Poisson}
\end{equation}

In our implementation, we normally set $N_p=10$, which results in less than $1\%$ probability of empty spots according to Eq.~\ref{Eq.Poisson}. While also a standard interpolation could be performed, a moving average approach is more robust against noise. The velocity corresponding to each bin $\boldsymbol{u}(x,y)$ is then the weighted average of the velocities of the particles falling in it. The scattered vector $u_{p_i}$ can be computed by super-resolution particle tracking, but without leading generality, any other kind of technique for feature matching can be exploited. The weight coefficients $c_i$ are set as a function of the distance $d_i$ of the $i^{th}$ particle from the bin centre: 

\begin{equation}
\begin{aligned}
   \boldsymbol{u}(x,y)  &= \frac{\sum\limits_{i = 1}^{N_p}{\boldsymbol{u_{p_{i}}} c_{i}} }{\sum\limits_{i = 1}^{N_p}{c_i}}, \,\,\,\,\,\,
   c_{i}  &= e ^ {-\left(\frac{d_{i}}{b_{ref}/2}\right)^2}. 
   \end{aligned}
   \label{Eq.weightedAverageRef}
\end{equation}

This weighting system has been adapted from the work of \citet{agui1987performance}. 

\vspace{0.1cm}
\textbf{Step 3: Local Proper Orthogonal Decomposition} 
A metric for the similarity between different snapshots must be set. While enforcing  it on the full snapshot appears overly restrictive, it can be argued that it could be feasible at a local level. To this purpose, a domain partition in subdomains is set, each one containing $N_v$ vectors. This value is related to the dimensions of the subdomain: at a first glance reducing the size of the subdomain opens the possibility to exploit more targeted features, but on the other hand extreme reduction makes difficult the correct evaluation of neighbours, which are all very similar due to the reduced rank of the snapshot matrix. In the remainder of this work, without leading generality, we employ squared subdomains, resembling interrogation windows in PIV processes also in terms of size.

Similarity between subdomains in different snapshots is assessed in low-order coordinates to reduce the dimension of the search space and to guarantee robustness to noise. To this purpose, POD is performed on each subdomain over the ensemble of realisations. Only the most energetic modes are retained to describe the feature space, i.e. only the first $r$ temporal modes.

Each subdomain, or window, is treated as a realisation and all the time samples for each subdomain are employed to build a snapshot matrix $\boldsymbol{U_{w}}$, of size $N_t \times (N_vN_c)$, with $N_t$ being the total number of snapshots, and $N_c$ the number of velocity components. A Singular Value Decomposition leads to:

\begin{equation}
    \boldsymbol{U_{w}} =  \boldsymbol{\Psi_{w}\Sigma_{w}\Phi_{w}^T}.
    \label{Eq.LocalPOD}
\end{equation}

The matrices $\boldsymbol{\Psi_{w}}$ and $\boldsymbol{\Phi_{w}}$ include the temporal and spatial modes, while $\boldsymbol{\Sigma_{w}}$ contains the singular values. It must be remarked that the operation in Eq.~\ref{Eq.LocalPOD} is performed on the fluctuating velocity fields after subtracting the average of ensemble PTV velocity fields at $b_{ref}$. This centring operation is necessary to avoid that the KNN search is strongly biased by the first mode being coincident with the average flow field.

%\begin{figure}[htb]
%\centering
%\includegraphics[scale=0.5]{3-windows.png}
%\includegraphics[scale=0.5]{3-Csum.png}
%\caption{Cumulative energy distribution in the highlighted windows: the first one is in the wake of the cylinder, the second one outside (a). The chosen rank is the number of modes that collects the 90\% of energy (b). }
%\label{Fig.csum}
%\centering
%\end{figure}

The features for KNN search are built by $\boldsymbol{\Theta_{w}}=\boldsymbol{\Psi_{w}}\boldsymbol{\Sigma_{w}}$ for each specific subdomain, truncated at rank $r$, which is set here as the number of modes collecting $90\%$ of energy. This rule of thumb is in qualitative agreement with the elbow criterion for the cases studied in this work. %Fig.\ref{Fig.csum} reports the example of fluidic pinball, described in next section, where two window's energy content are compared.

The KNN  algorithm searches the closest neighbours in the dataset according to its features $\boldsymbol{\theta_{w_i}}=\boldsymbol{\psi_{w_i}\sigma_{w_i}}$, corresponding to the POD temporal coefficients at the $i^{th}$ time instant. The process is repeated for each subdomain of the snapshot. 

\vspace{0.1cm}
\textbf{Step 4: Optimal-$K$ computation} 

For each instantaneous realisation of a subdomain, merging vectors from the $K$ similar snapshots leads to dense clouds of vectors. This operation is equivalent to increase locally the particle image density by a factor $K$. The bin size can thus be reduced accounting to this fictitious increase of $N_{ppp}$, leading to: 

\begin{equation}
   b_{HR} = \sqrt{\frac{N_{p}}{N_{ppp}K}}. 
    \label{Eq.btg}
\end{equation}

In our implementation $N_p$ is still fixed to $10$ (following classical rules of thumb from PIV theory). 

The choice of $K$ is then relevant for the size of the bin and for the final accuracy of the prediction. Low $K$ values lead to moderate bin size reduction (i.e. limited increase of spatial resolution); on the other hand, high $K$ values require extending the search neighbourhood, thus increasing the risk of merging snapshots with lower similarity in the process. Ideally, for $K=N_t$ the procedure converges to EPTV with bins containing on average $N_p$ vectors. The criterion for the selection of $K$ is based on minimising the dispersion of vector values within the bin. The optimal value of $K$ is obtained by spanning $K$ from $1$, corresponding to $b_{ref}$ (Eq.~\ref{Eq.bref}), to the one that allows a bin size of $4$ pixel. The $K$ providing minimum uncertainty is then selected. To speed up the process, the investigation is carried out on a limited number of snapshots and only on a few $K$ values. The trend of the uncertainty on $K$  is obtained by a cubic interpolation on the entire span, and the value providing minimum uncertainty on the interpolated curve is selected. We checked that normally $5$ to $10$ values of $K$ are enough to obtain the curve with sufficient accuracy. We discuss in the following step that this is equivalent to search for $arg\,min_K \,\hat{\delta}$, with $\hat{\delta}$ being the velocity uncertainty.  

Obtaining this map of $K$ can represent for large datasets the bulk of the computational cost of the entire procedure. As we will show in Sec.~\ref{sec: exp validation}, imposing symmetries or exploiting statistical homogeneity can provide a shortcut. A discussion on the time complexity of the process is reported in the Appendix.

\begin{table*}[t]
  \begin{flushleft}
  \rule{\textwidth}{0.04pt}
  Input: set of $N_t$ particle image pairs.
  \rule{\textwidth}{0.04pt}
  \begin{enumerate}
  
      \item \textit{Perform PTV analysis on original images.}
      \item  \textit{Compute the reference bin  (Eq.~\ref{Eq.bref}) and weighting of particles falling inside  (Eq.~\ref{Eq.weightedAverageRef}}) to obtain the binned reference distribution.
      \item \textit{Decompose by local POD (Eq.~\ref{Eq.LocalPOD}) and build local features dataset $\boldsymbol{\Theta_{w}}=\boldsymbol{\Psi_{w}}\boldsymbol{\Sigma_{w}}$ }, truncated at rank $r$.
      \item \textit{Search the  $arg\,min_K \,\hat{\delta}$ and from this the corresponding $b_{HR}$ via Eq.~\ref{Eq.btg}}.
      \item \textit{Perform KNN, rebuild the snapshots as in Eq.~\ref{Eq.wi} and compute the uncertainty $\hat{\delta}$ from Eq.~\ref{Eq.wstd}. } 
  \end{enumerate}
  \rule{\textwidth}{0.04pt}
  Output: high-resolution flow field and uncertainty estimation.
  \rule{\textwidth}{0.04pt}
  \caption{Algorithm of the KNN-PTV technique.} 
  \label{tab:algorithm}
    \end{flushleft}
\end{table*}

\vspace{0.1cm}
\textbf{Step 5: Velocity field reconstruction and uncertainty estimation}

On the basis of the selected $K$, and corresponding bin size from Eq.~\ref{Eq.btg}, the velocity for each bin is evaluated as a weighted average of the contributions of all vectors falling in the bin for the $K$ neighbours. Differently from Step 2, here vectors belonging to different snapshots are used. The weights thus take into account both the distance $d_i$ of the vector to the bin centre and the distance between the neighbours in the feature space $d_f$.

\begin{equation}
   \begin{aligned}
   \boldsymbol{u}(x,y)  &= \frac{\sum\limits_{i = 1}^{N_p}{\boldsymbol{u_{p_{i}}} c_{i}} }{\sum\limits_{i = 1}^{N_p}{c_i}}, \,\,\,\,\,\,
   c_{i} = e ^ {-\left(\frac{d_{i}}{\frac{b_{HR}}{2}}\right)^2}e ^ {-\left(\alpha \frac{d_{f_{jk}}}{\| \theta_{w_{z_j}} \|}\right)^2}.  
   \end{aligned}
   \label{Eq.wi}
\end{equation}

The parameter $d_{f_{jk}} $ is the euclidean distance in the feature space between the $k^{th}$ snapshot (where the $i^{th}$ particle falls) and the $j^{th}$ snapshot (the  one to rebuild), divided by the norm of $\theta_{w_{z_j}}$ that represents the $z^{th}$ feature test window (the one that contains the bin that contains the $i^{th}$ particle), evaluated only on the $j^{th}$ raw (the snapshot to rebuild).  The coefficient $\alpha$ is set empirically equal to $4$ to give more relevance to the spatial distance with respect to the distance in the feature space. This value, as shown in Sec.~\ref{sec: numerical validation}-\ref{sec: exp validation}, fits well for very different test cases, so we can assume, with a good confidence level, that is not flow-dependent. It must be remarked that in Eq.~\ref{Eq.wi} the $u_{p_{i}}$ represent the original velocity vectors, mean flow included.

Interestingly enough, the statistical dispersion of the vectors in Eq.~\ref{Eq.wi} can be exploited as an indicator of uncertainty, in analogy to the particle disparity method by \cite{sciacchitano2013piv}. In our case, the disparity vectors are set as the difference between the value assigned by Eq.~\ref{Eq.wi} and the velocity vector used for its evaluation. The instantaneous uncertainty $\hat{\delta}$ is defined as the weighted standard deviation of the disparity vectors:

\begin{equation}
   \hat{\delta} = \sqrt{\frac{\sum\limits_{i = 1}^{N_p} c_i\left(\boldsymbol{u_{p_{i}}}-\frac{\sum\limits_{i = 1}^{N_p} {(c_i\boldsymbol{u_{p_{i}}}})}{\sum\limits_{i=1}^{N_p}{c_i}}\right)^2}{\frac{M-1}{M} \sum\limits_{i = 1}^{N_p} c_i}},
   \label{Eq.wstd}
\end{equation}

\noindent where $M$ is the number of nonzero weights. The expanded uncertainty $U$ is related to $\hat{\delta}$ according to the definition given in \citet{coleman2018experimentation}:

\begin{equation}
     U = t \, \hat{\delta},
     \label{Eq:expandedUncertainty}
\end{equation}
\noindent where the factor $t$ is a coverage factor that comes from the tabulation of T-Student distribution, employed to approximate the $\hat{\delta}$ distribution. Expanded uncertainty allows associating a level of confidence to the estimation, or in other words, bounds containing the true value within a certain probability.

\vspace{0.2cm}
Step 4 of the process leverages the uncertainty minimisation according to Eq.~\ref{Eq.wstd} to select an optimal $K$ for each subdomain. The KNN-PTV algorithm pushes towards exploiting neighbours until increasing $K$ leads to a growing dispersion of the vectors (i.e. uncertainty). In this sense the algorithm is  ``adaptive'', i.e. it locally selects an optimal bin size according to uncertainty minimisation. This operation could, in principle, be carried out on each snapshot individually, thus leading to an optimal $K$-map for each instantaneous realisation. In practice, this is computationally intensive. Furthermore, as will be discussed in the validation, the sensitivity of the uncertainty to $K$ is rather low. For this reason, the approach adopted here is to establish a fixed $K$ map computed on statistical grounds on a limited number of snapshots.

\section{Validation}
\label{sec: numerical validation}
\subsection{Fluidic pinball}
\label{subsec: pinball}
\subsubsection{Database and numerical settings} \label{subsubsec: databpin}

The first test case is the flow around and in the wake of a fluidic pinball \citep{deng2020low}. Three cylinders with equal diameter $D$ are allocated to form with their centres an equilateral triangle with a side length equal to $3D/2$. The downstream side is  located at $x/D = 0$, centred with respect to the $y$ axis and it is oriented orthogonal to the freestream flow.

Despite its relative simplicity, this configuration is characterised by a complex flow behaviour, which exhibits chaos for a sufficiently large Reynolds number. Direct Numerical Simulation (DNS) data at $Re = 130$ are produced to generate a synthetic test case with a total of $4737$ snapshots. The flow field extends between $x/D\in[ -5D, 19D]$ in the streamwise direction and $y/D \in [-4D, 4D]$ in the crosswise direction.  The field is discretised with a resolution of  $ 25~\mathrm{pixels}/D$.

Synthetic PTV results are generated in this domain considering a random particle distribution with a particle image density $N_{ppp}$ of $0.02$ particles per pixel. The output high-resolution grid is set with a spacing of $4$ pixels leading to $48\times144$ vectors.
Particle images are generated for standard PIV analysis, with Gaussian-shaped particles having a maximum intensity of $100$ counts and a diameter of $2$ pixels.
 
In this test case, the KNN-PTV is fed with the exact position of the particles. This approach does not account for random noise due to particle positioning. We decided to adopt it in this first stage of validation as an attempt to isolate the error due to blending snapshots from other sources related to the image pairing process.

\subsubsection{High-resolution fields}\label{subsubsec: HRpin}

Following Fig.~\ref{Fig.flow}, the first step is to build  the reference low-resolution distribution to perform the local POD analysis. We impose that each bin contains $10$ particles, thus from the Eq.~\ref{Eq.bref} $b_{ref}=23$ pixels. The local POD is performed on subdomains containing $10\times10$ vectors (i.e. $40\times40$ pixels since the high-resolution grid is set with a spacing of $4$ pixels) with an overlap of approximately $75\%$.

\begin{figure}[t]
\centering
\begin{overpic}[scale=1,unit=1mm]{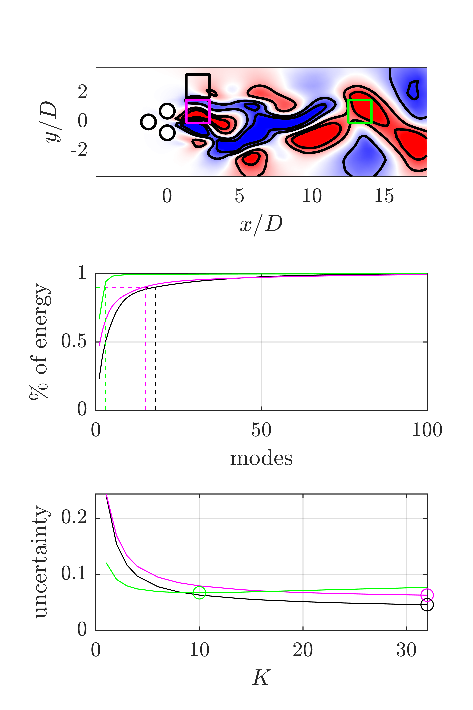}
\put(10,40){\textnormal{c)}}
\put(10,77){\textnormal{b)}}
\put(10,110){\textnormal{a)}}
\end{overpic}
\caption{Contour representation of the instantaneous fluctuating streamwise velocity field (a) and cumulative energy distribution of the local POD (b) in selected regions: immediately downstream of the pinball but outside of the wake (black); near wake of the pinball (magenta); far field (green). The regions are indicated with colour-coded squares in (a). Dashed lines in (b) indicate the selected rank to collects the $90\%$ of energy. Estimated mean uncertainty of velocity vectors for different values of $K$ according to Eq.~\ref{Eq.wstd} (c).}
\label{Fig.csum}
\centering
\end{figure}

For each subdomain, we compute the corresponding snapshot matrices and perform the POD analysis. Figure \ref{Fig.csum} shows the cumulative energy plot for three different windows: two windows in the near field, respectively outside and inside the wake, and a third window located in the far field. The regions are identified in Fig.~\ref{Fig.csum}a. Dashed lines in Fig.~\ref{Fig.csum}b indicate the number of modes needed to include $90\%$ of the energy in the feature definition process. The selection of the dimension of the feature space is locally adapted. For the windows outside of the wake, the fluctuation energy is spread on a larger number of modes. On the other hand, for the windows in the wake the majority of the energy information is collected by the first modes. This is particularly evident for the far field, where the large-scale vortex shedding becomes the dominating feature. 

With the selected rank for each window, the optimal $K$ selection is carried out, as reported in Fig.~\ref{Fig.csum}c. This process consists in spanning $K$ for each subdomain to identify the $K$ minimising the weighted standard deviation of the disparity vectors from Eq.~\ref{Eq.wstd}. It can be expected that an increase in the number of neighbours $K$ (thus increasing the particle image density with particles from local-neighbour snapshots) leads to reduced modulation effects, as it allows using smaller bin size. On the other hand, higher $K$ also implies merging velocity vectors from ``farther'' (in the feature space) snapshots, which increases the random error due to non-perfect correspondence in the flow structure. The maximum $K$ is set to $32$, corresponding to a minimum bin size of approximately $4$ pixels according to Eq.~\ref{Eq.btg}. We observe that in the near field the minimum of uncertainty is obtained using a larger number of neighbours, with the algorithm pushing towards the maximum allowed value of $K$. In this region the modulation effect is dominant, thus explaining the trend towards large $K$, and correspondingly small $b_{HR}$. On the other hand, in the far field we observe the opposite situation. In this region the modulation effects are less relevant, thus the minimum of uncertainty is achieved with a smaller value of $K$ and consequently higher $b_{HR}$. Interestingly, the curves of estimated uncertainty against $K$ exhibit a sort of \textit{plateau}. This hints towards the possibility of obtaining a good guess of a range of sub-optimal $K$ with very few attempts. 

\begin{figure}[t]
\centering
%\includegraphics[scale=0.7]{figs/2-Kop.png}
%\vfill
%\includegraphics[scale=0.7]{figs/2-btg.png}
\includegraphics[scale=1]{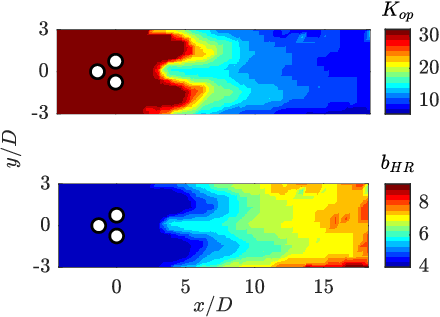}
\caption{Map of the optimal $K$ (top) and the corresponding map of bin size according to Eq.~\ref{Eq.btg} in pixels (bottom).}
\label{Fig.MAPPINB}
\end{figure}

The optimal $K$ is estimated for each subdomain. For all bins contained within a subdomain we use the same value of $K$. Following the criteria described in the previous section, we obtain the maps shown in Fig.~\ref{Fig.MAPPINB}. For this test case, the domain can be ideally split in three different regions. In the upstream region it is possible to reduce the bin size thanks to the availability of more neighbours very similar to each other since the flow exhibits limited variability. As we move further from the near field  downstream of the cylinder the flow becomes more chaotic, reducing the similarity between the nearest neighbours. Furthermore, as modulation effects become less important, the random error introduced by small differences between neighbours arises. Both effects reduce the possibility of including neighbours, thus not allowing to push towards smaller bin size.

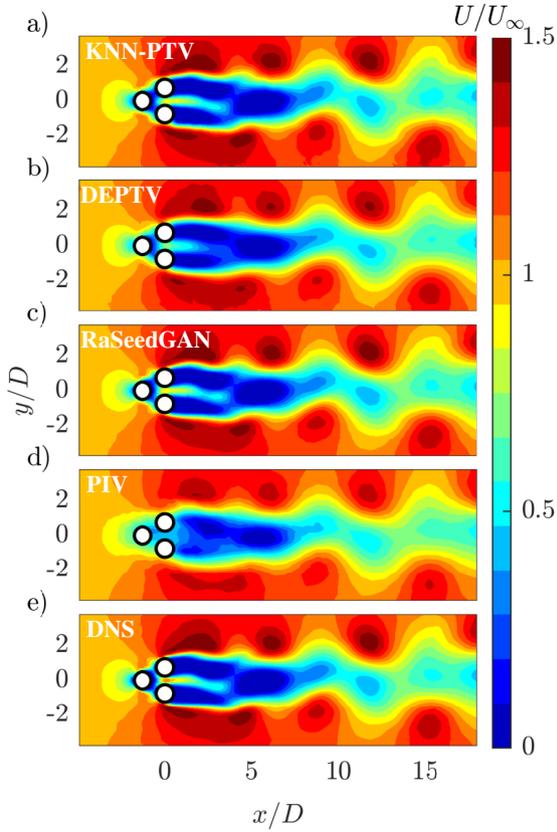
\begin{figure}[t]
\centering
\begin{overpic}[scale=1,unit=1mm]{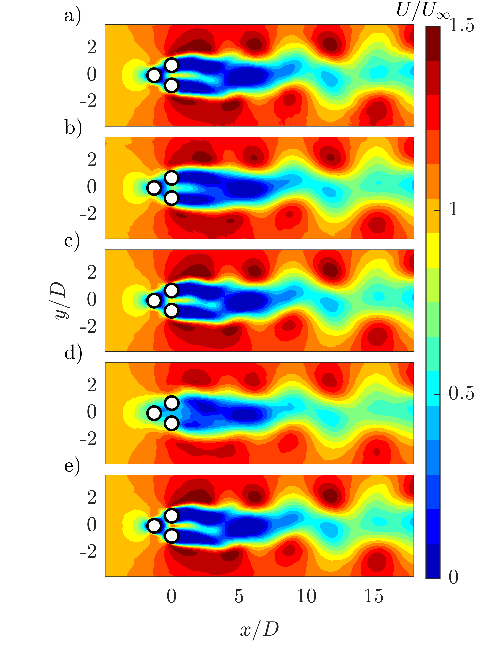}
\put(18,27){\parbox{10mm}{\textcolor{white}{\textnormal{ \textbf{DNS}}}}}
\put(18,46){\parbox{10mm}{\textcolor{white}{\textnormal{ \textbf{PIV}}}}}
\put(18,66){\parbox{10mm}{\textcolor{white}{\textnormal{ \textbf{RaSeedGAN}}}}}
\put(18,85){\parbox{10mm}{\textcolor{white}{\textnormal{ \textbf{DEPTV}}}}}
\put(18,103){\parbox{15mm}{\textcolor{white}{\textnormal{ \textbf{KNN-PTV}}}}}
\end{overpic}
\caption{Instantaneous streamwise velocity field contours estimated with: a) KNN-PTV, b) DEPTV, c) RaSeedGAN, d) standard PIV with interrogation window of $32\times32$ pixels. The reference field from the original DNS is included for comparison (e).}
\label{Fig.CompPinball}
\end{figure}

Figure \ref{Fig.CompPinball} includes a qualitative comparison between the proposed KNN-PTV, DEPTV \citep{cortina2021sparse} and RaSeedGAN \citep{guemes2022super}, all performed on the same number of snapshots. The DNS used to generate the images is used as ground truth. A standard cross-correlation multi-pass image-deformation-based analysis with an interrogation window of $32\times32$ pixels is included for reference. RaSeedGAN and KNN-PTV provide the best results with very small differences. DEPTV is also able to recover correctly the peaks of the streamwise velocity in the released wake. In all three cases there is a clear improvement in spatial resolution with respect to the reference cross-correlation analysis with window deformation.

\subsubsection{Reconstruction error}\label{subsubsec: pinerror}

In this section, we analyse quantitatively the reconstruction error. The adopted metric is the normalised root mean square error $\delta_{RMS}$, evaluated as:

\begin{equation}\label{eq.rms}
   \delta_{RMS} = \frac{\sqrt{\frac{\sum\limits_{i = 1}^{N_t}(U_i-U_{DNS_i})^2 + (V_i-V_{DNS_i})^2}{N_t}} }{U_\infty},
\end{equation}

\noindent where $U$ and $V$ are the estimated velocity vectors, $U_{DNS}$ and $V_{DNS}$ are the corresponding vectors from DNS described in Sec.~\ref{subsubsec: databpin}, $N_t$ is the number of snapshots and $U_\infty$ is the freestream velocity, the parameter chosen to obtain a dimensionless estimation. For our choice of time separation between frames, the displacement corresponding to the bulk velocity is equal to $1$ pixel. This error is a function of the position; with a little abuse of notation, we do not include explicitly this dependence.

Table \ref{Tab.ERRPin} reports the parameter $\langle\delta_{RMS}\rangle$, i.e. a spatial mean of the above mentioned $\delta_{RMS}$. For DEPTV and RaSeedGAN the test was performed with different bin sizes but for brevity in the table only the best results are reported. The KNN-PTV is carried out on datasets with a different number of images ($1000$,  $2000$, $3000$ and $4737$). 

Although RaSeedGAN obtains the best performance, we observe that KNN-PTV is able to progressively reduce the error by increasing the number of samples.  

\begin{table}[htb]
\centering
 \begin{tabular}{||c|c|c| c||} 
 \hline
 Method & Nt & $\langle\delta_{RMS}\rangle$ & bin size [pixels] \\ [0.5ex] 
 \hline\hline
 KNN-PTV &  4737 & 0.0299 & adaptive\\ 
 KNN-PTV &  3000 & 0.0308 & adaptive\\ 
 KNN-PTV & 2000& 0.0316 & adaptive\\ 
 KNN-PTV & 1000& 0.0337 & adaptive\\ 
 DEPTV &4737 & 0.0481 & 16 \\
 RaSeedGAN &4737 & 0.0191 & 4 \\
 PIV  &4737 & 0.0830 & 32 \\ [1ex] 
% PIV Nppp = 0.05 & 0.0602 & IW = 16 \\ [1ex] 
 \hline
 \end{tabular}
 \caption{Spatial average of the root mean square error $\langle\delta_{RMS}\rangle$, evaluated with Eq.~\ref{eq.rms}, for different methods.}
\label{Tab.ERRPin}
\end{table}

Finally, we attempt to pinpoint the sources of error in the KNN-PTV process. According to the Guide to Expression of Uncertainty in Measurement of the International Organisation for Standardisation \citep{schneider2017international} the errors can be categorised as random and systematic. While the first one is related to aleatory processes that can affect a measurement, the second one, according to the definition accepted by the PIV community, might not be constant, as reported in \citet{sciacchitano2019uncertainty}, and in contrast with the definition provided by \citet{coleman2018experimentation}. Following  \citet{scarano2003theory}, in cross-correlation-based PIV analysis (and in general for algorithms estimating instantaneous/average velocity fields as a moving average, see e.g. \citealp{kahler2012resolution,raiola2020adaptive}) the systematic error is proportional to the second-order spatial derivative of the velocity field and to the window size. To describe this effect, we search evidences of correlation between the error distribution and the hessian tensor $\boldsymbol{H}$, defined as:

\begin{equation}
 \boldsymbol{H} =    
 \begin{bmatrix}
\boldsymbol{u}_{xx} & \boldsymbol{u}_{xy} \\
\boldsymbol{u}_{xy} & \boldsymbol{u}_{yy}
\end{bmatrix}
.
\end{equation}

Since it is based on moving averages (although both in spatial and feature-space dimensions), a systematic error source can be identified in the modulation effect due to the second-order spatial derivative. Our results show a relatively weak, but still significant correlation, with the mean correlation coefficient between $\boldsymbol{\delta}= \sqrt{\boldsymbol{u}^2 - \boldsymbol{u}_{DNS}^2}$ and the Frobenius norm of $\boldsymbol{H}$ being $0.434$.

%\begin{equation}
%\boldsymbol{\delta} = \sqrt{\boldsymbol{u}^2 - %\boldsymbol{u}_{DNS}^2} 
%\end{equation}

On the other hand, the random error can be ascribed to the random distribution and the finite number of particles involved in the weight averaging process. It is to be expected a relation with the magnitude of the instantaneous velocity gradient within the averaging bin. Indeed, it must be remarked that the vectors obtained by KNN-PTV are the results of a weighted average, where the vectors belonging to the snapshot in question have higher weights if compared to particles extracted from neighbouring snapshots. Considering that, especially for small bin size, only a small portion of the vectors building the average pertains to the snapshot to be reconstructed, this can bias the estimation in case of presence of intense velocity gradients. We identified a significant correlation between error and velocity gradient, being the correlation coefficient equal to $0.573$. Clearly it must be remarked that other sources of error are present, e.g. the difference in instantaneous velocity of neighbouring snapshots if not sufficiently similar, while other error sources could not be modelled by this simulated experiment (positioning particle errors, outliers, etc.).

\subsubsection{Uncertainty validation}\label{subsubsec: uncpin}

In this subsection, we assess the weighted standard deviation of the disparity vectors from Eq.~\ref{Eq.wstd} as a metric to estimate the uncertainty. The approach follows the disparity method proposed by \cite{sciacchitano2013piv}, with the significant difference that the vectors are here weighted according to their distance from the bin centre and from the snapshot in the feature space. 

As reported by \citet{timmins2012method}, an easy way is the evaluation of  the ``uncertainty effectiveness'', which corresponds to the expanded uncertainty explained in Sec.~\ref{sec: algorithm description}. In this case the exact number of particles involved in the computation of velocity for each bin has been taken into account, allowing the evaluation of the exact coverage factor $t$ for each of them in the case of $95\%$ confidence from the T-Student tabulation \citep{coleman2018experimentation}. For this test case the value of uncertainty effectiveness computed on both the velocity components and averaged over $100$ snapshots is $90\%$, i.e. the uncertainty estimated by KNN-PTV is slightly non-conservative. 

\begin{figure}[t]
\centering
\begin{overpic}[scale=1,unit=1mm]{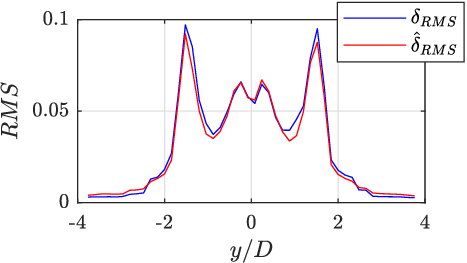}
\put(13.5,35.5){\textnormal{$\rho = 0.995$}}
\end{overpic}
\caption{Comparison between  $\delta_{RMS}$ and $\hat{\delta}_{RMS}$ at  x/D  = 1. In the top-left corner of the figure the correlation coefficient $\rho$ is also reported.}
\label{Fig.UncPin}
\end{figure}

Following \citet{sciacchitano2013piv}, the assessment of the uncertainty estimation is carried out also from a statistical perspective in time and space. A comparison is carried out between the real root mean square error, $\delta_{RMS}$, versus the estimated one, $\hat{\delta}_{RMS}$, on a fixed streamwise position corresponding to $x/D  = 1$. Fig.~\ref{Fig.UncPin} shows that the estimated uncertainty profile follows very closely the real error with a correlation coefficient $\rho = 0.994$, supporting the meaningfulness of the proposed uncertainty quantification method. 

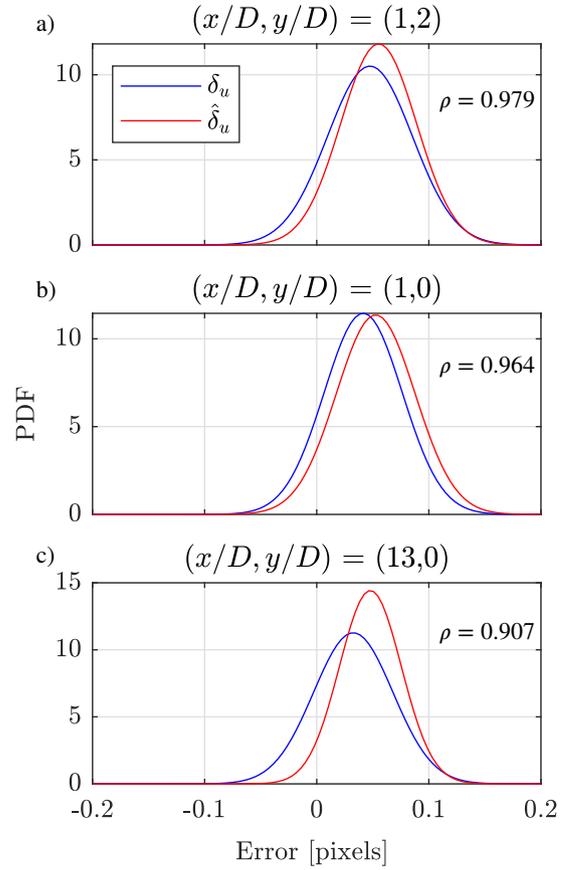
\begin{figure}[t]
\centering
\begin{overpic}[scale=1,unit=1mm]{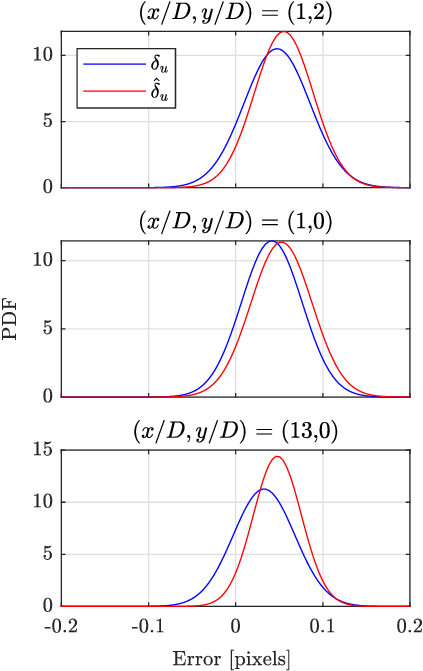}
\put(3,40){\textnormal{c)}}
\put(3,75){\textnormal{b)}} 
\put(3,110){\textnormal{a)}} 
\put(56,100){\textnormal{$\rho = 0.979 $}}
\put(56,65){\textnormal{$\rho = 0.964 $}} 
\put(56,30){\textnormal{$\rho = 0.907 $}} 
\end{overpic}
\caption{PDF of the instantaneous error $\delta_{u}$ and $\hat{\delta_{u}}$  for three different points: a) x/D = 1 and y/D = 2, b) x/D = 1 and y/D = 0 and c) x/D = 13 and y/D = 0. For each one the correlation coefficient $\rho$ is also reported.  }
\label{Fig.PDFPin}
\end{figure}

We also analyse the PDFs of the instantaneous error $\delta_{u}$ at three different locations, clarified in Fig.~\ref{Fig.PDFPin}. Also in this case the PDFs of the estimated error are very close to the real ones: the correlation range from about $91\%$ to $98\%$. The small discrepancy might be ascribed to the reduced systematic error due to the larger scale size in the far field.

\subsection{Channel Flow}

\subsubsection{Database and numerical settings}
\label{subsubsec:channel}
The second synthetic test case is based on a DNS of a turbulent channel flow, available at the Johns Hopkins Turbulence Database (\url {http://turbulence.pha.jhu.edu/}). This test case is expected to be more challenging, owing to the more chaotic flow behaviour and the difficulty in identifying locally-similar snapshots. 

The dimensions of the channel are $2$ half-channel-heights $h$ from wall to wall, $3\pi h$ in the span-wise direction and $8\pi h$ in the stream-wise direction. The reader is referred to \citet{li2008public} for all the details of the simulation settings.
For this simulated experiment, we extract subdomains of $2h\times h$ in the streamwise and wall-normal directions, respectively. The resolution is $512~\mathrm{pixels}/h$ and the particle image density is set to $0.01$ particles per pixel. The snapshots are generated with a time separation of $1$ convective time to reduce the correlation between different samples. Flow homogeneity in the streamwise and spanwise directions is exploited to extract a large number of snapshots. The separation between subdomains is $2h$ in the streamwise and $0.25h$ in the spanwise direction. The total number of generated snapshots is $11856$. Following the same approach of the pinball test case, PIV velocity fields with an interrogation window of $32\times32$ are generated. 

\subsubsection{Measurement error}\label{subsubsec: HRCha}

\begin{figure}[t]
\centering
\includegraphics[scale=1]{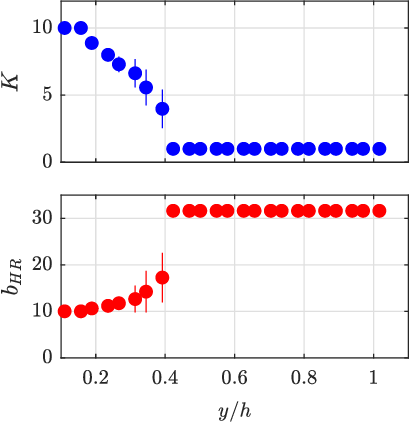}
\caption{Profile of the optimal $K$, on the top, and of the corresponding  bin size, $b_{HR}$, on the bottom. The values of $K$ and $b_{HR}$ are represented by circles (blue and red respectively) and evaluated as the mean along the streamwise direction. The vertical lines represent the standard deviation of these values.  }
\label{Fig.MAPCHA}
\centering
\end{figure}

The first step of the process is the computation of the optimal number of neighbours $K$, and the corresponding bin size, for each vector. The results are shown in Fig.~\ref{Fig.MAPCHA}. The profiles of $K$ (blue circle) and $b_{HR}$ (red circle) are computed as mean of the values on the map along the streamwise direction. In addition,  the standard deviations of these values are reported as vertical lines. The algorithm selects higher $K$ values near the wall, thus allowing locally to increase the resolution. This result can seem at first glance at odds with the intuition that higher local variance might induce more variability in the sample dataset, thus rendering it difficult to identify sufficiently close neighbours. The reason is to be sought in the analysis of the error sources. A larger number of neighbours (thus smaller bin size) allows systematic reduction of bias errors due to finite spatial resolution on the mean field, while on the other hand increases the random error due to dissimilarities between the identified neighbours. The trade-off between these two sources of error occur at larger $K$ in regions where the systematic error on the mean flow is expected to be higher, i.e. in the near-wall region. 

\begin{figure}[t]
\centering
\begin{overpic}[scale=1,unit=1mm]{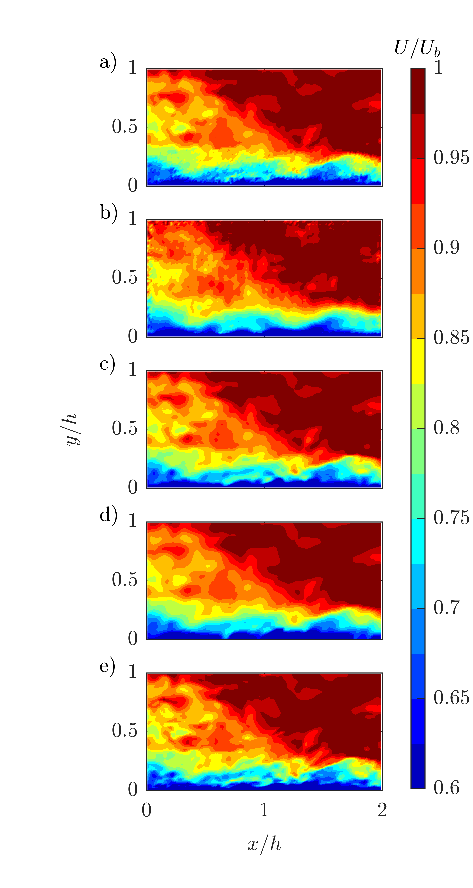}
\put(25,34){\parbox{10mm}{\textcolor{white}{\textnormal{ \textbf{DNS}}}}}
\put(24.5,59.5){\parbox{10mm}{\textcolor{white}{\textnormal{ \textbf{PIV}}}}}
\put(25,86){\parbox{10mm}{\textcolor{white}{\textnormal{ \textbf{RaSeedGAN}}}}}
\put(25,111.5){\parbox{10mm}{\textcolor{white}{\textnormal{ \textbf{DEPTV}}}}}
\put(25,136){\parbox{15mm}{\textcolor{white}{\textnormal{ \textbf{KNN-PTV}}}}}

\end{overpic}
\caption{Comparison of the same snapshot from: a) KNN-PTV, b) DEPTV, c) RaSeedGAN, d) PIV with interrogation window of $32\times32$ and e) the  reference DNS.}
\label{Fig.CompCha}
\end{figure}
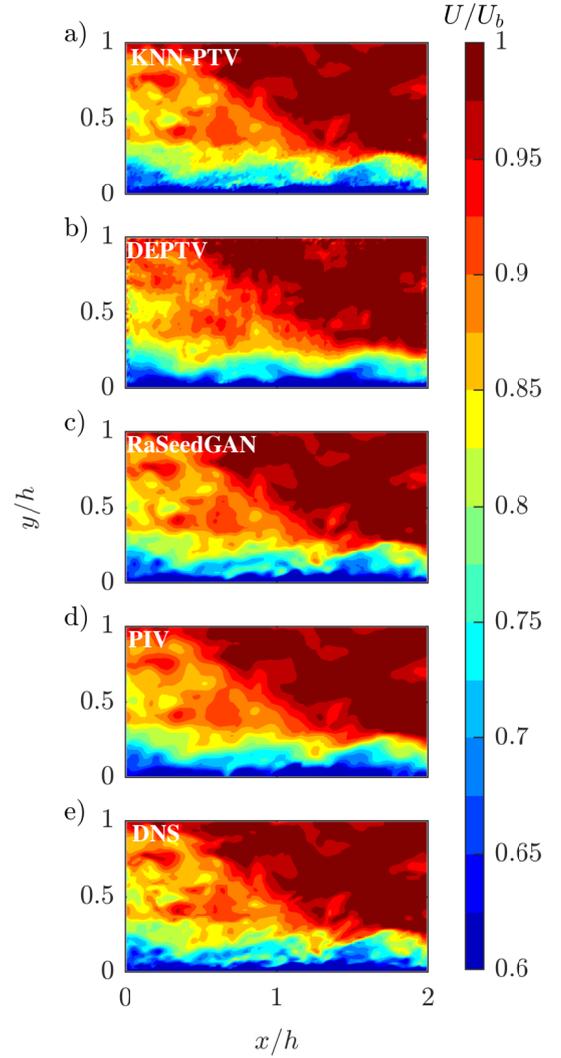

A qualitative comparison is reported in Fig.~\ref{Fig.CompCha}, while a quantitative comparison is included in Tab.~\ref{Tab.ERRCHA} in terms of  $\langle\delta_{RMS}\rangle$, computed as in Eq.~\ref{eq.rms} (in which $U_\infty$ is equal to 7.5 pixels). Among the tested techniques, DEPTV seems the least accurate. This result is not surprising, since DEPTV is expected to work best for flows with relatively-compact POD spectrum, e.g. free shear flows. RaSeedGAN is the most accurate method for this dataset. KNN-PTV highlights only slightly lower performances. It must be remarked, nonetheless, that for this dataset the performances of KNN-PTV are expected to improve very slightly with increasing dataset size since Fig.~\ref{Fig.MAPCHA} reports an optimal $K=1$ for $y/h >0.4$. Also for RaSeedGAN such improvement is not observed unless the neural-network architecture is made heavier \citep{guemes2022super}. 

\begin{table}[htb]
\centering
 \begin{tabular}{||c|c| c| c||} 
 \hline
 Method & Nt & $\langle\delta_{RMS}\rangle$  & bin size [pixels]  \\ [0.5ex] 
 \hline\hline
 KNN-PTV  & 11856 & 0.0207 & adaptive\\ 
% KNN-PTV  & 8000 & XXXX & adaptive\\
% KNN-PTV  & 5000 & XXXX & adaptive\\
 %KNN-PTV  & 1000 & XXXX & adaptive\\
 DEPTV & 11856 & 0.0398 & 16 \\
 RaSeedGAN & 11856 &0.0171 & 4 \\
 PIV  &11856 & 0.0248 & 32 \\ [1ex] 
 \hline
 \end{tabular}
 \caption{Spatial average of the root mean square error $\langle\delta_{RMS}\rangle$ evaluated for different methods: KNN-PTV, DEPTV, RaSeedGAN, PIV $32x32$.     }
\label{Tab.ERRCHA}
\end{table}

\subsubsection{Uncertainty validation}\label{subsubsec: unccha}

\begin{figure}[t]
    \centering
    \includegraphics[scale = 1]{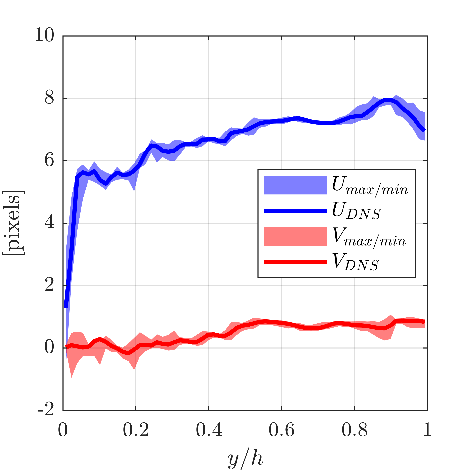}
    \caption{Instantaneous velocity profiles (continuous lines) of the streamwise and wall-normal velocity component at $x/h=1$. Transparent areas indicate the uncertainty bands with $95$\% confidence interval.}
    \label{Fig.bandscha}
\end{figure}

The uncertainty validation is carried out similarly to Sec.~\ref{subsubsec: uncpin}. The uncertainty effectiveness in this case has been measured to be $95\%$, in excellent agreement with the theoretical value. Figure \ref{Fig.bandscha} shows the uncertainty bands corresponding to an instantaneous realisation and both velocity components at $x/h=1$. The uncertainty bands are computed as: 

\begin{equation}
\begin{aligned}
 U_{max/min}   = U_{KNN} \pm t_{95}\hat{\delta}_u\\
 V_{max/min}   = V_{KNN} \pm t_{95}\hat{\delta}_v,
\end{aligned}
\end{equation}

\noindent where $t_{95}$ is the coverage factor described in Sec.~\ref{sec: algorithm description} for a  confidence level of $95\%$.

\begin{figure}[t]
\centering
\begin{overpic}[scale=1,unit=1mm]{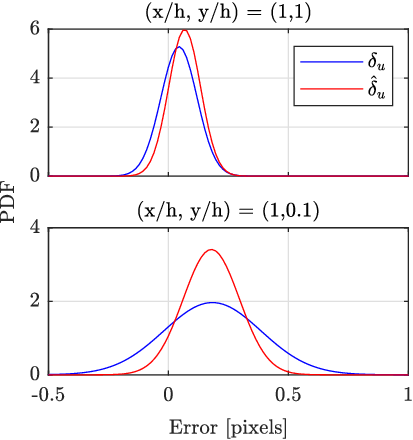}
\put(15,65){\textnormal{$\rho = 0.964 $}}
\put(15,30){\textnormal{$\rho = 0.942 $}} 
\end{overpic}
\caption{PDF of the instantaneous error $\delta_{u}$ and $\hat{\delta_{u}}$  for two different points: on the top $(x/h,y/h) = (1,1)$, on the bottom $(x/h,y/h) = (1,0.1)$. The correlation coefficient $\rho$ between estimated uncertainty and error distribution is included.}
\label{Fig.PDFCha}
\end{figure}

The statistical assessment in the time domain is shown for $(x/h, y/h) = (1,1)$ and $(x/h, y/h) = (1,0.1)$ in Fig.~\ref{Fig.PDFCha}. In the region far from the wall the estimation of the uncertainty is quite accurate, with a high degree of correlation with the statistical distribution of errors. In the near-wall region the PDFs are wider, as expected due to the higher intensity of the velocity fluctuations and stronger velocity gradients. In this case the agreement is slightly weaker. This can be observed also in Fig.~\ref{Fig.bandscha}, where an instantaneous profile of the streamwise and wall-normal velocity is included with the corresponding uncertainty bands.

\section{Experimental validation: turbulent boundary layer}
\label{sec: exp validation}
%\subsection{Experimental setup}\label{subsec: exp setup}
The algorithm is also validated with experimental data from \cite{guemes2019experimental}. The experiments are carried out in a G\"ottingen-type wind tunnel with test section length of $1.5 m$  and cross-sectional area of $0.4 \times 0.4$  $m^{2}$. The freestream turbulence intensity is below $1\%$ for velocities up to $20$ m/s.

\begin{figure}[t]
    \centering
  \begin{overpic}[width=1\columnwidth,unit=1mm]{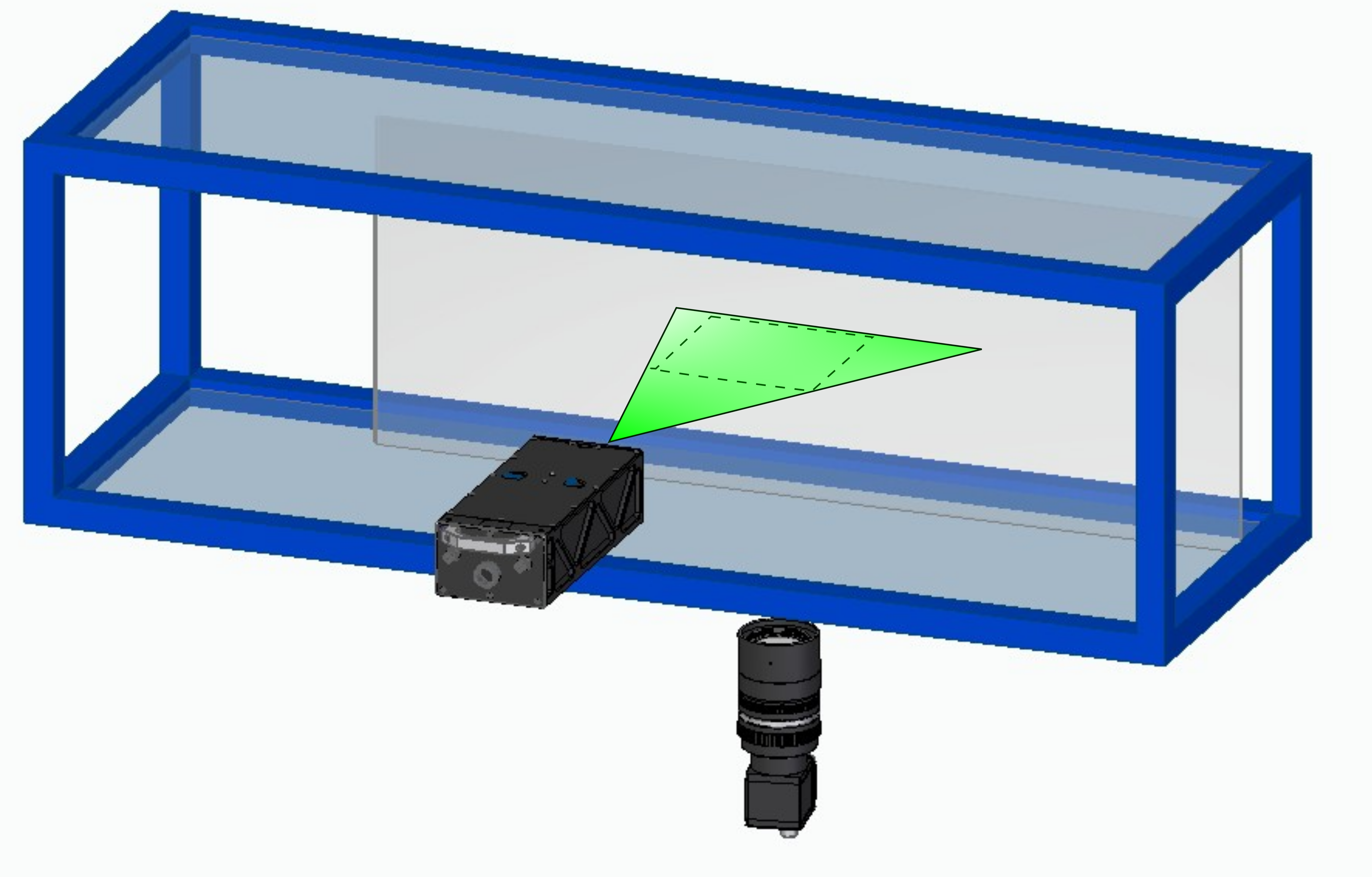}
  \put(15,36){\color{black}\vector(10,-1){10}}
  \put(15,37){\parbox{10mm}{\centering $U_{\infty}$}}
  \put(28,12){\circled{1}}
  \put(51,8){\circled{2}}
  \put(60,28){\circled{3}}
  \put(44,31){\circled{4}}
  %\put(5,35){ \circled{5}}
  \end{overpic}

    \caption{Sketch of the experimental setup. (1) Ng:Yag Quantel Evergreen laser; (2) ANDOR Zyla sCMOS $5.5$ MP camera; (3) methacrylate flat plate; (4) field of view of camera.  }
    \label{fig.expsetup}
\end{figure}
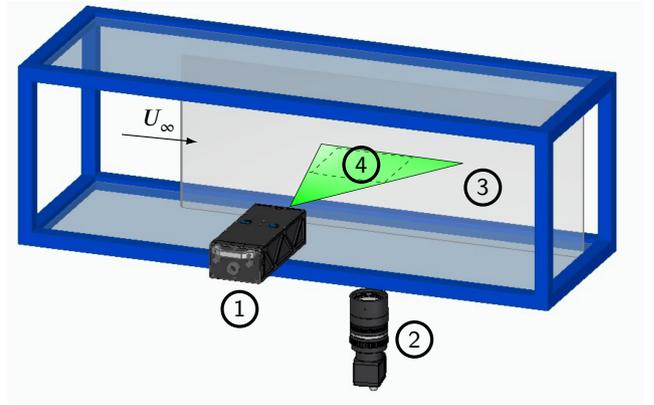

A turbulent boundary layer develops on a smooth methacrylate flat plate, as sketched in Fig.~\ref{fig.expsetup}. Droplets of Di-Ethyl-Hexyl-Sebacate (DEHS) with $1$ $\mu m$ diameter were employed to seed the flow. A light sheet produced by a dual cavity Ng:Yag Quantel Evergreen laser ($200$ mJ/pulse at $10$ Hz) provided the illumination. The images were recorded by an ANDOR Zyla sCMOS $5.5$ MP camera ($2560 \times 2160$ pixel array, $6.5 \times 6.5$ $\mu$m pixel size) with a resolution of about $48.5$ pix/mm.
All the details about the experiment are reported in \citet{guemes2019experimental}.

%\subsubsection{High resolution field}\label{subsubsec: HRTBL}

Velocity vectors are extracted with a super-resolution PTV approach \citep{keane1995super}. A multi-step image deformation algorithm \citep{scarano2001iterative} based on high-accuracy interpolation schemes \citep{astarita2005analysis,astarita2007analysis} is used to determine the predictor for the biased search. 

The PTV analysis delivers on average $25000$ vectors per snapshots, corresponding to a $0.006$ vectors per pixel. In order to establish a ground truth with higher accuracy, we randomly eliminate vectors from the dataset, reducing the number of vectors to $2560$ and consequently the density by a factor $10$. This corresponds in average to $10$ particles in a bin of $128 \times 128$ pixels.   The starting point for the building of binned matrices is a PIV with an interrogation window of $128 \times 128$ and overlap of $25\%$, the ground-truth will be a PIV with an interrogation window of $32 \times 32$ and an overlap of $25\%$. The total amount of snapshots employed is $30000$.

\begin{figure}[t]
    \centering
    \includegraphics [scale = 1]{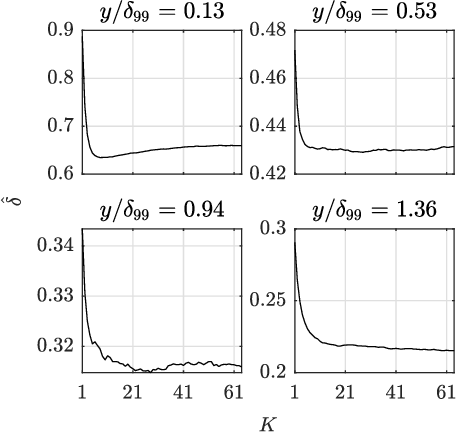}
    \caption{Uncertainty curves for windows centred at $x/\delta_{99} = 1$ and at four different $y/\delta_{99}$: $0.13$, $0.53$, $0.94$ and $1.36$.}
    \label{fig:plateaux}
\end{figure}

Owing to the large number of snapshots and grid points, the computational cost for estimating a full map of $K$ might be intimidating at first glance. As in the channel synthetic test case, statistical homogeneity in the streamwise direction can be enforced, thus requiring only to compute a profile. Nonetheless, we show that the sensitivity to the selection of $K$ is very small, thus paving the way to an estimation of the optimal-$K$ with very few points. An analysis of the uncertainty behaviour with $K$ on a large portion of the dataset ($10000$ snapshots) for a finite set of grid positions is illustrated in Fig.~\ref{fig:plateaux}. A \textit{plateau} is observed in all cases, as mentioned in Sec.~\ref{subsubsec: HRpin}. This can be ascribed to the limited amount of information brought by additional neighbours when their distance in the feature space becomes larger. Once the minimum number of $K$ to reach the plateau is identified (in all the tested grid points of the order of $10-20$), the selection can be extended to the rest of the domain. Paying the price of a slight increase of error, the computational costs is significantly reduced.  

\begin{figure}[t]
\centering
\begin{overpic}[scale=1,unit=1mm]{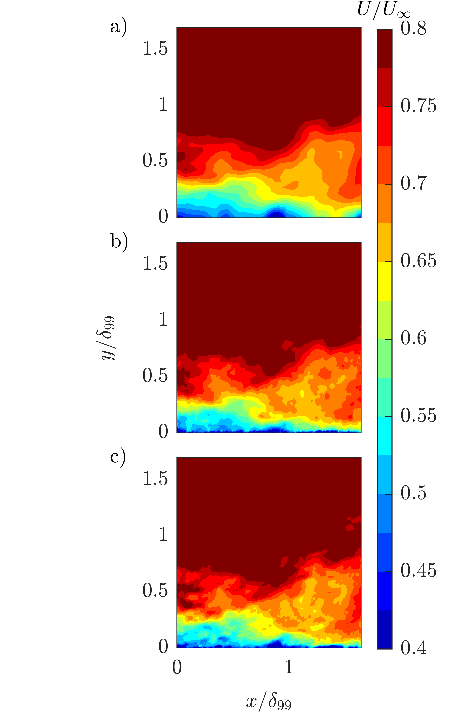}
\put(30,41){\parbox{15mm}{\textcolor{white}{\textnormal{ \textbf{PIV32}}}}}
\put(30,78){\parbox{15mm}{\textcolor{white}{\textnormal{ \textbf{KNN-PTV}}}}}
\put(30,114.5){\parbox{15mm}{\textcolor{white}{\textnormal{ \textbf{PIV128}}}}}
\end{overpic}
\caption{Comparison between: a) PIV with interrogation window of $128\times128$ pixels, b) KNN-PTV reconstruction, c) PIV with interrogation window of $32\times32$ pixels. The parameter $\delta_{99}$ for the dimensionless coordinates is the boundary layer thickness.}
\label{Fig.CompEXp}
\end{figure}

In Fig.~\ref{Fig.CompEXp} a qualitative comparison of the streamwise velocity fields obtained with KNN-PTV and the original PIV analysis is reported. The results are presented in non-dimensional form using the boundary layer thickness $\delta_{99}$ and the freestream velocity $U_\infty$, equal to $24.7$ mm and $15.5$ m/s, respectively. The high-resolution target (i.e. PIV with interrogation window of $32 \times 32$ pixels) is included for reference. KNN-PTV is able to recover small scales that are filtered out by the original PIV analysis with $128 \times 128$ pixels interrogation window. The non-dimensional error $\langle\delta_{RMS}\rangle$ for KNN-PTV is $0.0170$, while for PIV analysis with interrogation window of $128 \times 128$ pixels is equal to $0.0186$.

\begin{figure}[t]
\centering
\begin{overpic}[scale=1,unit=1mm]{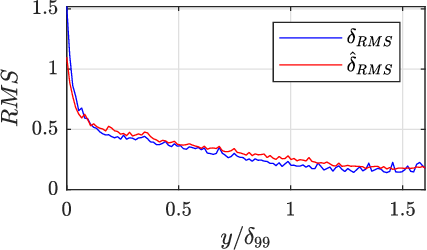}
\put(15,35){\textnormal{$\rho = 0.971 $}}
\end{overpic}
\caption{Comparison between  $\delta_{RMS}$ and $\hat{\delta}_{RMS}$ at  $x/\delta_{99}  = 1$.}
\label{Fig.UncChaEXP}
\end{figure}

%\subsubsection{Uncertainty validation}\label{subsubsec: uncTBL}
We validate the uncertainty estimation with the same procedure performed in Sec.~\ref{subsubsec: uncpin}-\ref{subsubsec: unccha}. 
In this case, the uncertainty effectiveness is $92\%$ against the theoretical $95\%$, which is again a satisfactory agreement. Figure \ref{Fig.UncChaEXP} shows a comparison between the profiles along the wall-normal direction of $\delta_{RMS}$ and $\hat{\delta}_{RMS}$. The two curves show excellent agreement, with a correlation coefficient equal to $\rho$ = $0.959$. As expected, higher uncertainty are in the near-wall region due to the strongest velocity gradients. 

\begin{figure}[t]
\centering
\begin{overpic}[scale = 1,unit = 1mm]{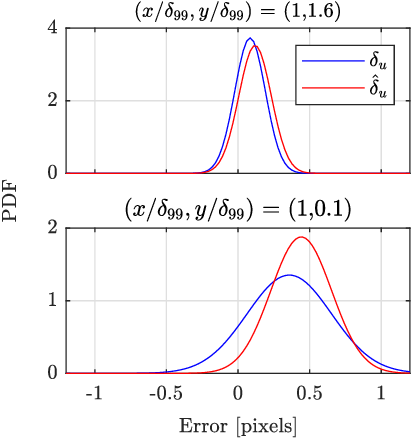}
\put(15,30){\textnormal{$\rho = 0.930 $}}
\put(15,65){\textnormal{$\rho = 0.970 $}}
\end{overpic}
\caption{PDF of the instantaneous error $\delta_{u}$ and $\hat{\delta_{u}}$  for two different points: on the top $x/D = 1$ and $y/D = 1.6$, on the bottom $x/D = 1$ and $y/D = 0.1$. For each one is reported also the correlation coefficient $\rho$. }
\label{Fig.PDFChaexp}
\end{figure}

The statistical distribution of real and estimated errors is performed at $x/\delta_{99} = 1$ and $y/\delta_{99} = (0.1,1.6)$, located respectively near and far the wall. The two PDFs in Fig.~\ref{Fig.PDFChaexp} confirm the trend of the synthetic case. There is a larger disagreement in the near-wall region, with a significant difference in systematic error. This can be ascribed in part by the use of experimental data as ``ground-truth''. Indeed, the reference data are also affected by finite spatial resolution in the near-wall region and by measurement noise.

\section{Conclusions}
A novel data-driven approach for the enhancement of PIV spatial resolution has been proposed.  The method is based on merging information (i.e. measured velocity vectors from PTV) of different non-time-resolved snapshots enforcing local similarity. The domain is split into subdomains, and for each of them a feature space is built to identify similar realisations. KNN is exploited to identify neighbouring subdomains in the feature space. The statistical dispersion of the velocity values of the vectors identified in each subdomain can also be exploited to estimate the uncertainty. Our method, named as KNN-PTV, provides an end-to-end tool for high-resolution measurements with uncertainty quantification directly embedded in the process. 

As proofed by the validation, KNN-PTV features the corresponding novelty and interesting properties:

\begin{itemize}
    \item Superior resolution if compared to standard PIV and techniques based on interpolating data from PTV. Theoretically, the resolution can be increased down to the sensitivity of particle positioning just by increasing the dataset size. In practice, a trade-off between dataset size and desired performance must be sought, similarly to EPTV methods.
    \item While DEPTV has shown remarkable performances for spectrally-compact flows, but poorer capabilities for flows with a wider richness of scales, KNN-PTV offers a robust resolution enhancement almost independently on the flow. In a way, KNN-PTV relaxes the linearity requirement imposed by DEPTV, which is global in time and space. With a local approach, and only using information from few significant neighbours per bin, KNN-PTV exploits all the advantages of the flexibility of a locally linear-embedding.
    \item KNN-PTV directly embeds uncertainty estimation in the process, which is a key feature for application and usage of measured data in models for uncertainty propagation.
    \item Even though it is slightly outperformed by more refined neural-network methods, KNN-PTV requires little if no expertise at all in the training phase. Once the vectors are available, the performances are weakly sensitive to the parameter choice, thus easing significantly its applicability.
\end{itemize}

In conclusion, KNN-PTV presents as a promising tool for high-resolution measurements with embedded uncertainty quantification. The proposed framework provides the flexibility to adapt locally the spatial resolution. KNN-PTV is conceived as an end-to-end tool, with a robust recipe and possibility to tune parameters with minimal expertise.  We foresee applications in the field of wall-bounded flows and, more in general, in moderate-to-high Reynolds number flows where time resolution is often not available. Future research efforts might be directed towards extension to volumetric measurements and/or inclusion of physical constraints to improve the accuracy.

\section*{Appendix - Time complexity discussion} 
\begin{table*}[htb]
\centering
 \begin{tabular}{||c| c||} 
 \hline
 Method &  Time complexity  \\ [0.5ex] 
 \hline\hline
 PIV  &  $\mathcal{O} \left(2 D_I^2log_2 \left(D_I\right)\frac{HW}{GD_{PIV}^2} N_{step}N_t + 
 4HW\left(2+v_{comp}\right) \left(N_{step}-1\right) N_{t}\right) $ \\
  \hline 
 KNN-PTV  &  $\mathcal{O}\left(d N_t log\left(N_t\right)N_tN_s + K log\left(N_t\right)N_tN_s + d_p n_p log\left(n_p\right) N_tN_s\right) $\\ 
  \hline
% KNN-PTV  & 8000 & XXXX & adaptive\\
% KNN-PTV  & 5000 & XXXX & adaptive\\
 %KNN-PTV  & 1000 & XXXX & adaptive\\
 DEPTV & $\mathcal{O}\left(N_p  N_{bin}  N_t + N_{vPIV} d N_t^2\right)$  \\
  \hline
 RaSeedGAN &  $\mathcal{O}\left( N_{t}\sum\limits_{l = 1}^{N_L} n_{l-1} s^2 n_{l} m_{l}^2\right)$ \\[1ex] 
 \hline
 \end{tabular}
 \caption{Time complexity evaluated for different methods: KNN-PTV, DEPTV, RaSeedGAN, PIV.     }
\label{Tab.time complexity}
\end{table*}
In this appendix, we estimate the time complexity of estimating a dataset of snapshots using KNN-PTV, DEPTV and RaSeedGAN. The time complexity is commonly evaluated by counting the number of elementary operations performed by the algorithm, supposing that each of them requires the same execution time. Nonetheless, we underline that actual computational cost differences depend upon implementation details and architecture. For instance, RaSeedGAN can count on efficient implementation on GPUs from TensorFlow, while KNN-PTV is released in Matlab, coded by the authors, without explicit use of parallelised computations.

First of all, we detail the workload of the different phases of the KNN-PTV process for the case of the fluidic pinball. The analysis has been carried out on a common laptop (CPU Intel Core i7-10750H   $2.60$ GHz, RAM $16$ GB) with the built-in Matlab profiler.

The snapshot reconstruction and the uncertainty estimation require most of the computational cost ($81\%$). The majority of this phase is spent on two main processes: KNN search ($39\%$) of neighbours local snapshots and construction of the trees for vector searches within each bin (kd-tree, $50\%$). For these operations, we can easily quantify the time complexity: for the kd-tree building is $\mathcal{O}(d N_tlog(N_t))$, where $d$ is the data dimensionality and $N_t$ is the population used to build the tree, i.e. the number of samples; on the other hand, the time complexity of solely the KNN search is $\mathcal{O}(K log(N_t))$. These estimations are performed following \citet{friedman1977algorithm}. \\Regarding the KNN search of neighbours, the Matlab function ``knnsearch'' builds automatically a kd-tree for the searching of neighbours when the dimensionality of the data is smaller than $10$. In this case the dimensionality is equal to the number of retained modes for feature search. The overall time complexity of the entire KNN process is $\mathcal{O}(d N_t log(N_t) + K log(N_t))$, where the dimensionality $d$ corresponds to the rank $r$, i.e. the number of retained modes. This amount of operations is repeated for each snapshot and each subdomain, so it is multiplied by $N_t$ and the number of subdomains $N_s$. 
\\
The time complexity of the kd-tree is $\mathcal{O}(d_p n_p log(n_p) N_t N_s)$, where $d_p=2$ is the dimensionality of the particle space, and the number of particles involved in the tree building is $n_p = N_{ppp}(W_s)^2K$, being $W_s$ the dimension of the subdomain, i.e. $40$ pixels. The rest of the processes are performed only once for the entire dataset: the computation of the K-map ($17\%$), the building of the reference distributions ($0.8\%$) and the assembling of the training set ($<0.1\%$). 
\\
The computation of the maps of $K$ is in principle intensive since it requires repeating the process above for a number of different $K$. In practice, in our experience this operation can be easily performed on a smaller dataset and testing only few $K$ values, since the map of $K$ and the curve of errors associated to its choice are normally smooth. 
The cost of building the reference binned distribution (that is the same for all the methodologies discussed in this appendix) is the time complexity of a ball query, $\mathcal{O}(N_p N_{bin} N_t)$, where $N_{bin}$ is the number of bins and $N_p$ is the number of particles within each image. The assembly of the local POD for neighbours identification is equal to the time complexity of a SVD performed on each window $\mathcal{O}((N_v d)^2 N_t N_s)$, where $N_v$ is the number of vectors of the subdomain, i.e. 100 vectors, and assuming that $(N_v d)<N_t$. 

Regarding DEPTV, the algorithm consists of 3 main steps: calculation of the reference POD modes from the low-resolution data; computation of the high-resolution ensemble POD modes; snapshot reconstruction. The computational complexity depends on the rank of the flow to be reconstructed. The first step of the process constitutes the bulk of the computational cost. In particular, there are two main processes within it: the building of the reference binned distribution and the SVD. Both of them have been already quantified in terms of time complexity: $\mathcal{O}(N_p  N_{bin}  N_t)$ for the first one and $\mathcal{O}(N_{vPIV} d N_t^2)$, assuming that $N_t>(N_{vPIV} d)$.

Regarding RaSeedGAN, the order of magnitude for time complexity can be quantified following \citet{he2015convolutional}:
\begin{equation}
    \mathcal{O}\left( N_t \sum\limits_{l = 1}^{N_L} n_{l-1} s_l^2 n_{l} m_{l}^2 \right),
    \label{eq.conv}
\end{equation}
where $l$ is the index of the convolutional layer, $N_L$ is the total number of layers, $n_{l-1}$ is the number of input channels in the $l^{th}$ layer, $n_{l}$ is the number of output channel in the $l^{th}$ layer, $s_l$ is the spatial size of the filter kernel and $m_l$ is the spatial size of the output feature map. We can model the generator and the discriminator as two convolutional networks with a certain number of layers as shown in Eq.~\ref{eq.conv}. However, this quantification is not straightforward, as it should also take into account the number of epochs for training.

For standard PIV it is possible to have an idea about the order of magnitude of its time complexity: 
\begin{equation}
\begin{split}
 \mathcal{O} \Biggl(2 D_I^2log_2 \left(D_I\right)\frac{HW}{GD_{PIV}^2} N_{step}N_t + \ldots  \\
 4HW\left(2+v_{comp}\right) \left(N_{step}-1\right) N_{t}\Biggl) ,
\end{split}
\end{equation}
where $D_I$ is the PIV interrogation window, $H$ and $W$ are the dimension in pixels of the input pictures, $GD_{PIV}$ is the PIV grid distance, $N_{step}$ is the number of iterations and $v_{comp}$ are the velocity components. The first term in parenthesis refers to the cross-correlation calculation with Fast Fourier Transform, and assuming for simplicity an interrogation window with constant size over the iterations. The second term refers instead to the interpolation process for image deformation, using for simplicity a bi-linear interpolation scheme.

 It is possible to quantify the time complexity of each method as ratio to the PIV value, as reported in Tab.~\ref{Tab.time complexity}. The analysis performed on the datasets described in this paper shows that the DEPTV  has an average time complexity of the same order as the PIV, instead for KNN-PTV is $\mathcal{O}(10-10^2)$ while for RaSeedGAN is $\mathcal{O}(10^3-10^4)$. We must warn the reader on this estimation for two main reasons. The first one is that the time complexity estimation is an order of magnitude analysis, which is suitable to investigate the scaling of each individual algorithm with processing parameters. Applying the above formulations for comparison of different algorithms is an exercise which must be treated with caution. Furthermore, these ratios cannot be linearly converted with the execution time of the algorithm because, as mentioned at the beginning of this appendix, it depends on several factors such as the programming language and implementation, the possibility of parallel operations, hardware availability, etc. For example, to rebuild the entire dataset of the fluidic pinball the KNN-PTV takes more or less 13 hours on a common laptop, while the RaSeedGAN, despite  performing a number of operations $100$ times greater than KNN-PTV, takes more or less 2-3 hours. This, as mentioned before, is due to the fact that the RaSeedGAN is coded on TensorFlow, which runs the code directly in GPU where parallel computations are optimised. Furthermore, we are not considering for instance overheads due to input/output of the algorithms, which are expected to have a more significant impact on the standard PIV analysis for being the algorithm with least time complexity of the set.

\vspace{0.1cm}
\section*{Acknowledgment}
This project has received funding from the European Research Council (ERC) under the European Union’s Horizon 2020 research and innovation program (grant agreement No 949085). Funding for APC: Universidad Carlos III de Madrid (Read \& Publish Agreement CRUE-CSIC 2022).
The authors warmly acknowledge N. Deng, B. Noack, M. Morzynski and L. Pastur for providing the code for the fluidic pinball simulations and A.G\"uemes for the experimental dataset.

\section*{Data availability}
All datasets used in this work are openly available in Zenodo, accessible at \url {https://doi.org/10.5281/zenodo.6922577}. All codes developed
in this work are openly available in GitHub, accessible through the
link: \url{https://github.com/erc-nextflow/KNN-PTV}.

%\appendix

%\section{Sample Appendix Section}

%\appendix
%\section{My Appendix}
%Appendix sections are coded under \verb+\appendix+.

%\verb+\printcredits+ command is used after appendix sections to list 
%author credit taxonomy contribution roles tagged using \verb+\credit+ 
%in frontmatter.

\printcredits

\bibliographystyle{cas-model2-names}

\bibliography{cas-refs}

\begin{thebibliography}{54}
\expandafter\ifx\csname natexlab\endcsname\relax\def\natexlab#1{#1}\fi
\providecommand{\url}[1]{\texttt{#1}}
\providecommand{\href}[2]{#2}
\providecommand{\path}[1]{#1}
\providecommand{\DOIprefix}{doi:}
\providecommand{\ArXivprefix}{arXiv:}
\providecommand{\URLprefix}{URL: }
\providecommand{\Pubmedprefix}{pmid:}
\providecommand{\doi}[1]{\href{http://dx.doi.org/#1}{\path{#1}}}
\providecommand{\Pubmed}[1]{\href{pmid:#1}{\path{#1}}}
\providecommand{\bibinfo}[2]{#2}
\ifx\xfnm\relax \def\xfnm[#1]{\unskip,\space#1}\fi
%Type = Article
\bibitem[{Adrian(1997)}]{adrian1997dynamic}
\bibinfo{author}{Adrian, R.}, \bibinfo{year}{1997}.
\newblock \bibinfo{title}{Dynamic ranges of velocity and spatial resolution of
  particle image velocimetry}.
\newblock \bibinfo{journal}{Measurement Science and Technology}
  \bibinfo{volume}{8}, \bibinfo{pages}{1393}.
%Type = Article
\bibitem[{Ag{\"u}era et~al.(2016)Ag{\"u}era, Cafiero, Astarita and
  Discetti}]{aguera2016ensemble}
\bibinfo{author}{Ag{\"u}era, N.}, \bibinfo{author}{Cafiero, G.},
  \bibinfo{author}{Astarita, T.}, \bibinfo{author}{Discetti, S.},
  \bibinfo{year}{2016}.
\newblock \bibinfo{title}{Ensemble 3d {PTV} for high resolution turbulent
  statistics}.
\newblock \bibinfo{journal}{Measurement Science and Technology}
  \bibinfo{volume}{27}, \bibinfo{pages}{124011}.
%Type = Article
\bibitem[{Ag{\"u}{\'\i} and Jimenez(1987)}]{agui1987performance}
\bibinfo{author}{Ag{\"u}{\'\i}, J.C.}, \bibinfo{author}{Jimenez, J.},
  \bibinfo{year}{1987}.
\newblock \bibinfo{title}{On the performance of particle tracking}.
\newblock \bibinfo{journal}{{Journal of Fluid Mechanics}}
  \bibinfo{volume}{185}, \bibinfo{pages}{447--468}.
%Type = Article
\bibitem[{Astarita(2007)}]{astarita2007analysis}
\bibinfo{author}{Astarita, T.}, \bibinfo{year}{2007}.
\newblock \bibinfo{title}{Analysis of weighting windows for image deformation
  methods in {PIV}}.
\newblock \bibinfo{journal}{Experiments in Fluids} \bibinfo{volume}{43},
  \bibinfo{pages}{859--872}.
%Type = Article
\bibitem[{Astarita(2009)}]{astarita2009adaptive}
\bibinfo{author}{Astarita, T.}, \bibinfo{year}{2009}.
\newblock \bibinfo{title}{Adaptive space resolution for {PIV}}.
\newblock \bibinfo{journal}{Experiments in Fluids} \bibinfo{volume}{46},
  \bibinfo{pages}{1115--1123}.
%Type = Article
\bibitem[{Astarita and Cardone(2005)}]{astarita2005analysis}
\bibinfo{author}{Astarita, T.}, \bibinfo{author}{Cardone, G.},
  \bibinfo{year}{2005}.
\newblock \bibinfo{title}{{Analysis of interpolation schemes for image
  deformation methods in PIV}}.
\newblock \bibinfo{journal}{Experiments in Fluids} \bibinfo{volume}{38},
  \bibinfo{pages}{233--243}.
%Type = Article
\bibitem[{Avallone et~al.(2015)Avallone, Discetti, Astarita and
  Cardone}]{avallone2015convergence}
\bibinfo{author}{Avallone, F.}, \bibinfo{author}{Discetti, S.},
  \bibinfo{author}{Astarita, T.}, \bibinfo{author}{Cardone, G.},
  \bibinfo{year}{2015}.
\newblock \bibinfo{title}{{Convergence enhancement of single-pixel PIV with
  symmetric double correlation}}.
\newblock \bibinfo{journal}{Experiments in Fluids} \bibinfo{volume}{56},
  \bibinfo{pages}{1--11}.
%Type = Article
\bibitem[{Beresh(2021)}]{beresh2021time}
\bibinfo{author}{Beresh, S.J.}, \bibinfo{year}{2021}.
\newblock \bibinfo{title}{Time-resolved particle image velocimetry}.
\newblock \bibinfo{journal}{Measurement Science and Technology}
  \bibinfo{volume}{32}, \bibinfo{pages}{102003}.
%Type = Article
\bibitem[{Cai et~al.(2019)Cai, Zhou, Xu and Gao}]{cai2019dense}
\bibinfo{author}{Cai, S.}, \bibinfo{author}{Zhou, S.}, \bibinfo{author}{Xu,
  C.}, \bibinfo{author}{Gao, Q.}, \bibinfo{year}{2019}.
\newblock \bibinfo{title}{Dense motion estimation of particle images via a
  convolutional neural network}.
\newblock \bibinfo{journal}{Experiments in Fluids} \bibinfo{volume}{60},
  \bibinfo{pages}{1--16}.
%Type = Article
\bibitem[{Cierpka et~al.(2013)Cierpka, L{\"u}tke and
  K{\"a}hler}]{cierpka2013higher}
\bibinfo{author}{Cierpka, C.}, \bibinfo{author}{L{\"u}tke, B.},
  \bibinfo{author}{K{\"a}hler, C.J.}, \bibinfo{year}{2013}.
\newblock \bibinfo{title}{Higher order multi-frame particle tracking
  velocimetry}.
\newblock \bibinfo{journal}{Experiments in Fluids} \bibinfo{volume}{54},
  \bibinfo{pages}{1--12}.
%Type = Book
\bibitem[{Coleman and Steele(2018)}]{coleman2018experimentation}
\bibinfo{author}{Coleman, H.W.}, \bibinfo{author}{Steele, W.G.},
  \bibinfo{year}{2018}.
\newblock \bibinfo{title}{{Experimentation, Validation, and Uncertainty
  Analysis for Engineers}}.
\newblock \bibinfo{publisher}{John Wiley \& Sons}.
%Type = Article
\bibitem[{Cortina-Fern{\'a}ndez et~al.(2021)Cortina-Fern{\'a}ndez, Vila, Ianiro
  and Discetti}]{cortina2021sparse}
\bibinfo{author}{Cortina-Fern{\'a}ndez, J.}, \bibinfo{author}{Vila, C.S.},
  \bibinfo{author}{Ianiro, A.}, \bibinfo{author}{Discetti, S.},
  \bibinfo{year}{2021}.
\newblock \bibinfo{title}{From sparse data to high-resolution fields: ensemble
  particle modes as a basis for high-resolution flow characterization}.
\newblock \bibinfo{journal}{Experimental Thermal and Fluid Science}
  \bibinfo{volume}{120}, \bibinfo{pages}{110178}.
%Type = Article
\bibitem[{Cowen and Monismith(1997)}]{cowen1997hybrid}
\bibinfo{author}{Cowen, E.}, \bibinfo{author}{Monismith, S.},
  \bibinfo{year}{1997}.
\newblock \bibinfo{title}{A hybrid digital particle tracking velocimetry
  technique}.
\newblock \bibinfo{journal}{Experiments in Fluids} \bibinfo{volume}{22},
  \bibinfo{pages}{199--211}.
%Type = Article
\bibitem[{Deng et~al.(2020)Deng, Noack, Morzy{\'n}ski and Pastur}]{deng2020low}
\bibinfo{author}{Deng, N.}, \bibinfo{author}{Noack, B.R.},
  \bibinfo{author}{Morzy{\'n}ski, M.}, \bibinfo{author}{Pastur, L.R.},
  \bibinfo{year}{2020}.
\newblock \bibinfo{title}{Low-order model for successive bifurcations of the
  fluidic pinball}.
\newblock \bibinfo{journal}{Journal of Fluid Mechanics} \bibinfo{volume}{884}.
%Type = Article
\bibitem[{Deng et~al.(2019)Deng, He, Liu and Kim}]{deng2019super}
\bibinfo{author}{Deng, Z.}, \bibinfo{author}{He, C.}, \bibinfo{author}{Liu,
  Y.}, \bibinfo{author}{Kim, K.C.}, \bibinfo{year}{2019}.
\newblock \bibinfo{title}{Super-resolution reconstruction of turbulent velocity
  fields using a generative adversarial network-based artificial intelligence
  framework}.
\newblock \bibinfo{journal}{Physics of Fluids} \bibinfo{volume}{31},
  \bibinfo{pages}{125111}.
%Type = Article
\bibitem[{Di~Florio et~al.(2002)Di~Florio, Di~Felice and
  Romano}]{di2002windowing}
\bibinfo{author}{Di~Florio, D.}, \bibinfo{author}{Di~Felice, F.},
  \bibinfo{author}{Romano, G.}, \bibinfo{year}{2002}.
\newblock \bibinfo{title}{Windowing, re-shaping and re-orientation
  interrogation windows in particle image velocimetry for the investigation of
  shear flows}.
\newblock \bibinfo{journal}{Measurement Science and Technology}
  \bibinfo{volume}{13}, \bibinfo{pages}{953}.
%Type = Article
\bibitem[{Discetti and Coletti(2018)}]{discetti2018volumetric}
\bibinfo{author}{Discetti, S.}, \bibinfo{author}{Coletti, F.},
  \bibinfo{year}{2018}.
\newblock \bibinfo{title}{Volumetric velocimetry for fluid flows}.
\newblock \bibinfo{journal}{Measurement Science and Technology}
  \bibinfo{volume}{29}, \bibinfo{pages}{042001}.
%Type = Article
\bibitem[{Fix and Hodges(1989)}]{fix1989discriminatory}
\bibinfo{author}{Fix, E.}, \bibinfo{author}{Hodges, J.L.},
  \bibinfo{year}{1989}.
\newblock \bibinfo{title}{Discriminatory analysis. nonparametric
  discrimination: Consistency properties}.
\newblock \bibinfo{journal}{International Statistical Review/Revue
  Internationale de Statistique} \bibinfo{volume}{57},
  \bibinfo{pages}{238--247}.
%Type = Article
\bibitem[{Friedman et~al.(1977)Friedman, Bentley and
  Finkel}]{friedman1977algorithm}
\bibinfo{author}{Friedman, J.H.}, \bibinfo{author}{Bentley, J.L.},
  \bibinfo{author}{Finkel, R.A.}, \bibinfo{year}{1977}.
\newblock \bibinfo{title}{An algorithm for finding best matches in logarithmic
  expected time}.
\newblock \bibinfo{journal}{ACM Transactions on Mathematical Software (TOMS)}
  \bibinfo{volume}{3}, \bibinfo{pages}{209--226}.
%Type = Article
\bibitem[{Gao et~al.(2021)Gao, Lin, Tu, Zhu, Wei, Zhang and
  Shao}]{gao2021robust}
\bibinfo{author}{Gao, Q.}, \bibinfo{author}{Lin, H.}, \bibinfo{author}{Tu, H.},
  \bibinfo{author}{Zhu, H.}, \bibinfo{author}{Wei, R.}, \bibinfo{author}{Zhang,
  G.}, \bibinfo{author}{Shao, X.}, \bibinfo{year}{2021}.
\newblock \bibinfo{title}{A robust single-pixel particle image velocimetry
  based on fully convolutional networks with cross-correlation embedded}.
\newblock \bibinfo{journal}{Physics of Fluids} \bibinfo{volume}{33},
  \bibinfo{pages}{127125}.
%Type = Article
\bibitem[{Goodfellow et~al.(2014)Goodfellow, Pouget-Abadie, Mirza, Xu,
  Warde-Farley, Ozair, Courville and Bengio}]{goodfellow2014generative}
\bibinfo{author}{Goodfellow, I.}, \bibinfo{author}{Pouget-Abadie, J.},
  \bibinfo{author}{Mirza, M.}, \bibinfo{author}{Xu, B.},
  \bibinfo{author}{Warde-Farley, D.}, \bibinfo{author}{Ozair, S.},
  \bibinfo{author}{Courville, A.}, \bibinfo{author}{Bengio, Y.},
  \bibinfo{year}{2014}.
\newblock \bibinfo{title}{Generative adversarial nets}.
\newblock \bibinfo{journal}{Advances in Neural Information Processing Systems}
  \bibinfo{volume}{27}.
%Type = Inproceedings
\bibitem[{G{\"u}emes et~al.(2019)G{\"u}emes, Ianiro and
  Discetti}]{guemes2019experimental}
\bibinfo{author}{G{\"u}emes, A.}, \bibinfo{author}{Ianiro, A.},
  \bibinfo{author}{Discetti, S.}, \bibinfo{year}{2019}.
\newblock \bibinfo{title}{Experimental assessment of large-scale motions in
  turbulent boundary layers}, in: \bibinfo{booktitle}{13th International
  Symposium on Particle Image Velocimetry}.
%Type = Article
\bibitem[{G{\"u}emes et~al.(2022)G{\"u}emes, Sanmiguel~Vila and
  Discetti}]{guemes2022super}
\bibinfo{author}{G{\"u}emes, A.}, \bibinfo{author}{Sanmiguel~Vila, C.},
  \bibinfo{author}{Discetti, S.}, \bibinfo{year}{2022}.
\newblock \bibinfo{title}{Super-resolution {GAN}s of randomly-seeded fields}.
\newblock \bibinfo{journal}{arXiv preprint arXiv:2202.11701} .
%Type = Article
\bibitem[{Hain and K{\"a}hler(2007)}]{hain2007fundamentals}
\bibinfo{author}{Hain, R.}, \bibinfo{author}{K{\"a}hler, C.},
  \bibinfo{year}{2007}.
\newblock \bibinfo{title}{Fundamentals of multiframe particle image velocimetry
  ({PIV})}.
\newblock \bibinfo{journal}{Experiments in Fluids} \bibinfo{volume}{42},
  \bibinfo{pages}{575--587}.
%Type = Inproceedings
\bibitem[{He and Sun(2015)}]{he2015convolutional}
\bibinfo{author}{He, K.}, \bibinfo{author}{Sun, J.}, \bibinfo{year}{2015}.
\newblock \bibinfo{title}{Convolutional neural networks at constrained time
  cost}, in: \bibinfo{booktitle}{Proceedings of the IEEE conference on computer
  vision and pattern recognition}, pp. \bibinfo{pages}{5353--5360}.
%Type = Article
\bibitem[{K{\"a}hler et~al.(2016)K{\"a}hler, Astarita, Vlachos, Sakakibara,
  Hain, Discetti, La~Foy and Cierpka}]{kahler2016main}
\bibinfo{author}{K{\"a}hler, C.J.}, \bibinfo{author}{Astarita, T.},
  \bibinfo{author}{Vlachos, P.P.}, \bibinfo{author}{Sakakibara, J.},
  \bibinfo{author}{Hain, R.}, \bibinfo{author}{Discetti, S.},
  \bibinfo{author}{La~Foy, R.}, \bibinfo{author}{Cierpka, C.},
  \bibinfo{year}{2016}.
\newblock \bibinfo{title}{{Main results of the 4th International PIV
  Challenge}}.
\newblock \bibinfo{journal}{Experiments in Fluids} \bibinfo{volume}{57},
  \bibinfo{pages}{1--71}.
%Type = Article
\bibitem[{K{\"a}hler et~al.(2012a)K{\"a}hler, Scharnowski and
  Cierpka}]{kahler2012resolution}
\bibinfo{author}{K{\"a}hler, C.J.}, \bibinfo{author}{Scharnowski, S.},
  \bibinfo{author}{Cierpka, C.}, \bibinfo{year}{2012}a.
\newblock \bibinfo{title}{On the resolution limit of digital particle image
  velocimetry}.
\newblock \bibinfo{journal}{Experiments in Fluids} \bibinfo{volume}{52},
  \bibinfo{pages}{1629--1639}.
%Type = Article
\bibitem[{K{\"a}hler et~al.(2012b)K{\"a}hler, Scharnowski and
  Cierpka}]{kahler2012uncertainty}
\bibinfo{author}{K{\"a}hler, C.J.}, \bibinfo{author}{Scharnowski, S.},
  \bibinfo{author}{Cierpka, C.}, \bibinfo{year}{2012}b.
\newblock \bibinfo{title}{On the uncertainty of digital piv and ptv near
  walls}.
\newblock \bibinfo{journal}{Experiments in Fluids} \bibinfo{volume}{52},
  \bibinfo{pages}{1641--1656}.
%Type = Article
\bibitem[{Keane et~al.(1995)Keane, Adrian and Zhang}]{keane1995super}
\bibinfo{author}{Keane, R.}, \bibinfo{author}{Adrian, R.},
  \bibinfo{author}{Zhang, Y.}, \bibinfo{year}{1995}.
\newblock \bibinfo{title}{Super-resolution particle imaging velocimetry}.
\newblock \bibinfo{journal}{Measurement Science and Technology}
  \bibinfo{volume}{6}, \bibinfo{pages}{754}.
%Type = Article
\bibitem[{Lagemann et~al.(2021)Lagemann, Lagemann, Mukherjee and
  Schr{\"o}der}]{lagemann2021deep}
\bibinfo{author}{Lagemann, C.}, \bibinfo{author}{Lagemann, K.},
  \bibinfo{author}{Mukherjee, S.}, \bibinfo{author}{Schr{\"o}der, W.},
  \bibinfo{year}{2021}.
\newblock \bibinfo{title}{Deep recurrent optical flow learning for particle
  image velocimetry data}.
\newblock \bibinfo{journal}{Nature Machine Intelligence} \bibinfo{volume}{3},
  \bibinfo{pages}{641--651}.
%Type = Inproceedings
\bibitem[{Ledig et~al.(2017)Ledig, Theis, Husz{\'a}r, Caballero, Cunningham,
  Acosta, Aitken, Tejani, Totz, Wang et~al.}]{ledig2017photo}
\bibinfo{author}{Ledig, C.}, \bibinfo{author}{Theis, L.},
  \bibinfo{author}{Husz{\'a}r, F.}, \bibinfo{author}{Caballero, J.},
  \bibinfo{author}{Cunningham, A.}, \bibinfo{author}{Acosta, A.},
  \bibinfo{author}{Aitken, A.}, \bibinfo{author}{Tejani, A.},
  \bibinfo{author}{Totz, J.}, \bibinfo{author}{Wang, Z.}, et~al.,
  \bibinfo{year}{2017}.
\newblock \bibinfo{title}{Photo-realistic single image super-resolution using a
  generative adversarial network}, in: \bibinfo{booktitle}{Proceedings of the
  IEEE Conference on Computer Vision and Pattern Recognition}, pp.
  \bibinfo{pages}{4681--4690}.
%Type = Article
\bibitem[{Li et~al.(2008)Li, Perlman, Wan, Yang, Meneveau, Burns, Chen, Szalay
  and Eyink}]{li2008public}
\bibinfo{author}{Li, Y.}, \bibinfo{author}{Perlman, E.}, \bibinfo{author}{Wan,
  M.}, \bibinfo{author}{Yang, Y.}, \bibinfo{author}{Meneveau, C.},
  \bibinfo{author}{Burns, R.}, \bibinfo{author}{Chen, S.},
  \bibinfo{author}{Szalay, A.}, \bibinfo{author}{Eyink, G.},
  \bibinfo{year}{2008}.
\newblock \bibinfo{title}{{A public turbulence database cluster and
  applications to study Lagrangian evolution of velocity increments in
  turbulence}}.
\newblock \bibinfo{journal}{Journal of Turbulence} , \bibinfo{pages}{N31}.
%Type = Article
\bibitem[{Lumley(1967)}]{lumley1967structure}
\bibinfo{author}{Lumley, J.L.}, \bibinfo{year}{1967}.
\newblock \bibinfo{title}{The structure of inhomogeneous turbulent flows}.
\newblock \bibinfo{journal}{Atmospheric turbulence and radio wave propagation}
  .
%Type = Article
\bibitem[{Lynch and Scarano(2013)}]{lynch2013high}
\bibinfo{author}{Lynch, K.}, \bibinfo{author}{Scarano, F.},
  \bibinfo{year}{2013}.
\newblock \bibinfo{title}{{A high-order time-accurate interrogation method for
  time-resolved PIV}}.
\newblock \bibinfo{journal}{Measurement Science and Technology}
  \bibinfo{volume}{24}, \bibinfo{pages}{035305}.
%Type = Article
\bibitem[{Nogueira et~al.(1999)Nogueira, Lecuona and
  Rodriguez}]{nogueira1999local}
\bibinfo{author}{Nogueira, J.}, \bibinfo{author}{Lecuona, A.},
  \bibinfo{author}{Rodriguez, P.}, \bibinfo{year}{1999}.
\newblock \bibinfo{title}{Local field correction {PIV}: on the increase of
  accuracy of digital {PIV} systems}.
\newblock \bibinfo{journal}{Experiments in Fluids} \bibinfo{volume}{27},
  \bibinfo{pages}{107--116}.
%Type = Article
\bibitem[{Novara et~al.(2012)Novara, Ianiro and Scarano}]{novara2012adaptive}
\bibinfo{author}{Novara, M.}, \bibinfo{author}{Ianiro, A.},
  \bibinfo{author}{Scarano, F.}, \bibinfo{year}{2012}.
\newblock \bibinfo{title}{{Adaptive interrogation for 3D-PIV}}.
\newblock \bibinfo{journal}{Measurement Science and Technology}
  \bibinfo{volume}{24}, \bibinfo{pages}{024012}.
%Type = Book
\bibitem[{Pope(2000)}]{pope2000turbulent}
\bibinfo{author}{Pope, S.B.}, \bibinfo{year}{2000}.
\newblock \bibinfo{title}{Turbulent flows}.
\newblock \bibinfo{publisher}{Cambridge University Press}.
%Type = Book
\bibitem[{Raffel et~al.(1998)Raffel, Willert, Kompenhans
  et~al.}]{raffel1998particle}
\bibinfo{author}{Raffel, M.}, \bibinfo{author}{Willert, C.E.},
  \bibinfo{author}{Kompenhans, J.}, et~al., \bibinfo{year}{1998}.
\newblock \bibinfo{title}{Particle image velocimetry: a practical guide}.
  volume~\bibinfo{volume}{2}.
\newblock \bibinfo{publisher}{Springer}.
%Type = Article
\bibitem[{Raiola et~al.(2020)Raiola, Lopez-Nu{\~n}ez, Cafiero and
  Discetti}]{raiola2020adaptive}
\bibinfo{author}{Raiola, M.}, \bibinfo{author}{Lopez-Nu{\~n}ez, E.},
  \bibinfo{author}{Cafiero, G.}, \bibinfo{author}{Discetti, S.},
  \bibinfo{year}{2020}.
\newblock \bibinfo{title}{{Adaptive ensemble PTV}}.
\newblock \bibinfo{journal}{Measurement Science and Technology}
  \bibinfo{volume}{31}, \bibinfo{pages}{085301}.
%Type = Article
\bibitem[{Sanmiguel~Vila et~al.(2017)Sanmiguel~Vila, {\"O}rl{\"u}, Vinuesa,
  Schlatter, Ianiro and Discetti}]{sanmiguel2017adverse}
\bibinfo{author}{Sanmiguel~Vila, C.}, \bibinfo{author}{{\"O}rl{\"u}, R.},
  \bibinfo{author}{Vinuesa, R.}, \bibinfo{author}{Schlatter, P.},
  \bibinfo{author}{Ianiro, A.}, \bibinfo{author}{Discetti, S.},
  \bibinfo{year}{2017}.
\newblock \bibinfo{title}{Adverse-pressure-gradient effects on turbulent
  boundary layers: statistics and flow-field organization}.
\newblock \bibinfo{journal}{Flow, Turbulence and Combustion}
  \bibinfo{volume}{99}, \bibinfo{pages}{589--612}.
%Type = Article
\bibitem[{Scarano(2001)}]{scarano2001iterative}
\bibinfo{author}{Scarano, F.}, \bibinfo{year}{2001}.
\newblock \bibinfo{title}{{Iterative image deformation methods in PIV}}.
\newblock \bibinfo{journal}{Measurement Science and Technology}
  \bibinfo{volume}{13}, \bibinfo{pages}{R1}.
%Type = Article
\bibitem[{Scarano(2003)}]{scarano2003theory}
\bibinfo{author}{Scarano, F.}, \bibinfo{year}{2003}.
\newblock \bibinfo{title}{Theory of non-isotropic spatial resolution in {PIV}}.
\newblock \bibinfo{journal}{Experiments in Fluids} \bibinfo{volume}{35},
  \bibinfo{pages}{268--277}.
%Type = Article
\bibitem[{Schanz et~al.(2016)Schanz, Gesemann and
  Schr{\"o}der}]{schanz2016shake}
\bibinfo{author}{Schanz, D.}, \bibinfo{author}{Gesemann, S.},
  \bibinfo{author}{Schr{\"o}der, A.}, \bibinfo{year}{2016}.
\newblock \bibinfo{title}{{Shake-The-Box: Lagrangian particle tracking at high
  particle image densities}}.
\newblock \bibinfo{journal}{Experiments in Fluids} \bibinfo{volume}{57},
  \bibinfo{pages}{1--27}.
%Type = Article
\bibitem[{Scharnowski et~al.(2012)Scharnowski, Hain and
  K{\"a}hler}]{scharnowski2012reynolds}
\bibinfo{author}{Scharnowski, S.}, \bibinfo{author}{Hain, R.},
  \bibinfo{author}{K{\"a}hler, C.J.}, \bibinfo{year}{2012}.
\newblock \bibinfo{title}{Reynolds stress estimation up to single-pixel
  resolution using {PIV}-measurements}.
\newblock \bibinfo{journal}{Experiments in Fluids} \bibinfo{volume}{52},
  \bibinfo{pages}{985--1002}.
%Type = Article
\bibitem[{Schneider et~al.(2017)Schneider, Maurer and
  Friedberg}]{schneider2017international}
\bibinfo{author}{Schneider, F.}, \bibinfo{author}{Maurer, C.},
  \bibinfo{author}{Friedberg, R.C.}, \bibinfo{year}{2017}.
\newblock \bibinfo{title}{{International organization for standardization (ISO)
  15189}}.
\newblock \bibinfo{journal}{Annals of Laboratory Medicine}
  \bibinfo{volume}{37}, \bibinfo{pages}{365--370}.
%Type = Article
\bibitem[{Sciacchitano(2019)}]{sciacchitano2019uncertainty}
\bibinfo{author}{Sciacchitano, A.}, \bibinfo{year}{2019}.
\newblock \bibinfo{title}{Uncertainty quantification in particle image
  velocimetry}.
\newblock \bibinfo{journal}{Measurement Science and Technology}
  \bibinfo{volume}{30}, \bibinfo{pages}{092001}.
%Type = Article
\bibitem[{Sciacchitano et~al.(2012)Sciacchitano, Scarano and
  Wieneke}]{sciacchitano2012multi}
\bibinfo{author}{Sciacchitano, A.}, \bibinfo{author}{Scarano, F.},
  \bibinfo{author}{Wieneke, B.}, \bibinfo{year}{2012}.
\newblock \bibinfo{title}{Multi-frame pyramid correlation for time-resolved
  {PIV}}.
\newblock \bibinfo{journal}{Experiments in Fluids} \bibinfo{volume}{53},
  \bibinfo{pages}{1087--1105}.
%Type = Article
\bibitem[{Sciacchitano et~al.(2013)Sciacchitano, Wieneke and
  Scarano}]{sciacchitano2013piv}
\bibinfo{author}{Sciacchitano, A.}, \bibinfo{author}{Wieneke, B.},
  \bibinfo{author}{Scarano, F.}, \bibinfo{year}{2013}.
\newblock \bibinfo{title}{{PIV} uncertainty quantification by image matching}.
\newblock \bibinfo{journal}{Measurement Science and Technology}
  \bibinfo{volume}{24}, \bibinfo{pages}{045302}.
%Type = Article
\bibitem[{Theunissen et~al.(2006)Theunissen, Scarano and
  Riethmuller}]{theunissen2006adaptive}
\bibinfo{author}{Theunissen, R.}, \bibinfo{author}{Scarano, F.},
  \bibinfo{author}{Riethmuller, M.L.}, \bibinfo{year}{2006}.
\newblock \bibinfo{title}{An adaptive sampling and windowing interrogation
  method in {PIV}}.
\newblock \bibinfo{journal}{Measurement Science and Technology}
  \bibinfo{volume}{18}, \bibinfo{pages}{275}.
%Type = Article
\bibitem[{Timmins et~al.(2012)Timmins, Wilson, Smith and
  Vlachos}]{timmins2012method}
\bibinfo{author}{Timmins, B.H.}, \bibinfo{author}{Wilson, B.W.},
  \bibinfo{author}{Smith, B.L.}, \bibinfo{author}{Vlachos, P.P.},
  \bibinfo{year}{2012}.
\newblock \bibinfo{title}{A method for automatic estimation of instantaneous
  local uncertainty in particle image velocimetry measurements}.
\newblock \bibinfo{journal}{Experiments in Fluids} \bibinfo{volume}{53},
  \bibinfo{pages}{1133--1147}.
%Type = Article
\bibitem[{Wang et~al.(2022a)Wang, Liu and Wang}]{wang2022dense}
\bibinfo{author}{Wang, H.}, \bibinfo{author}{Liu, Y.}, \bibinfo{author}{Wang,
  S.}, \bibinfo{year}{2022}a.
\newblock \bibinfo{title}{Dense velocity reconstruction from particle image
  velocimetry/particle tracking velocimetry using a physics-informed neural
  network}.
\newblock \bibinfo{journal}{Physics of Fluids} \bibinfo{volume}{34},
  \bibinfo{pages}{017116}.
%Type = Article
\bibitem[{Wang et~al.(2022b)Wang, Li, Liu, Wu, Hao, Zhang and
  He}]{wang2022deep}
\bibinfo{author}{Wang, Z.}, \bibinfo{author}{Li, X.}, \bibinfo{author}{Liu,
  L.}, \bibinfo{author}{Wu, X.}, \bibinfo{author}{Hao, P.},
  \bibinfo{author}{Zhang, X.}, \bibinfo{author}{He, F.}, \bibinfo{year}{2022}b.
\newblock \bibinfo{title}{Deep-learning-based super-resolution reconstruction
  of high-speed imaging in fluids}.
\newblock \bibinfo{journal}{Physics of Fluids} \bibinfo{volume}{34},
  \bibinfo{pages}{037107}.
%Type = Article
\bibitem[{Westerweel et~al.(2013)Westerweel, Elsinga and
  Adrian}]{westerweel2013particle}
\bibinfo{author}{Westerweel, J.}, \bibinfo{author}{Elsinga, G.E.},
  \bibinfo{author}{Adrian, R.J.}, \bibinfo{year}{2013}.
\newblock \bibinfo{title}{Particle image velocimetry for complex and turbulent
  flows}.
\newblock \bibinfo{journal}{Annual Review of Fluid Mechanics}
  \bibinfo{volume}{45}, \bibinfo{pages}{409--436}.
%Type = Article
\bibitem[{Westerweel et~al.(2004)Westerweel, Geelhoed and
  Lindken}]{westerweel2004single}
\bibinfo{author}{Westerweel, J.}, \bibinfo{author}{Geelhoed, P.},
  \bibinfo{author}{Lindken, R.}, \bibinfo{year}{2004}.
\newblock \bibinfo{title}{Single-pixel resolution ensemble correlation for
  micro-{PIV} applications}.
\newblock \bibinfo{journal}{Experiments in Fluids} \bibinfo{volume}{37},
  \bibinfo{pages}{375--384}.

\end{thebibliography}




\end{document}